\documentclass[fleqn,usenatbib]{mnras}

\usepackage{newtxtext,newtxmath}

\usepackage[T1]{fontenc}
\usepackage{ae,aecompl}

\usepackage{graphicx}
\usepackage{amsmath}
\usepackage{amssymb}
\usepackage[utf8]{inputenc}
\usepackage[version=4]{mhchem}
\usepackage{etoolbox}
\makeatletter
\patchcmd\@combinedblfloats{\box\@outputbox}{\unvbox\@outputbox}{}{%
  \errmessage{\noexpand\@combinedblfloats could not be patched}%
}%
\makeatother

\title[Chemical analysis of NGC\,6366]{Chemical analysis of eight giant stars of the globular cluster NGC\,6366}

\author[A. A. Puls et al.]{
Arthur A. Puls,$^{1}$\thanks{Based on observations collected at Canada France Hawaii Telescope and on public data available in the ESO Archive.}
\thanks
{E-mail: puls.arthur@gmail.com}
Alan Alves Brito,$^{1}$
Fabíola Campos,$^{2}$
Bruno Dias,$^{3}$\newauthor
and Beatriz Barbuy$^{4}$
\\
$^{1}$Departamento de Astronomia, Universidade Federal do Rio Grande do Sul, Av. Bento Gonçalves 9500,\\Porto Alegre 91501-970, RS, Brazil\\
$^{2}$Department of Astronomy, University of Texas at Austin, Austin, TX, USA\\
$^{3}$European Southern Observatory, Alonso de Cordova 3107, Santiago, Chile\\
$^{4}$Universidade de São Paulo, IAG, Rua do Matão 1226, Cidade Universitária, São Paulo 05508-900, Brazil
}

\date{Accepted 2018 January 29. Received 2018 January 26; in original form 2017 August 28}

\pubyear{2018}

\begin{document}
\label{firstpage}
\pagerange{\pageref{firstpage}--\pageref{lastpage}}
\maketitle

\begin{abstract}
The metal-rich Galactic globular cluster NGC\,6366 is the fifth closest to the Sun. Despite its interest, it has received scarce attention, and little is known about its internal structure. Its kinematics suggests a link to the halo, but its metallicity indicates otherwise.
We present a detailed chemical analysis of eight giant stars of NGC\,6366, using high resolution and high quality spectra (R $>$ 40\,000, S/N $>$ 60) obtained at the VLT (8.2~m) and CFHT (3.6~m) telescopes. We attempted to characterize its chemistry and to search for evidence of multiple stellar populations.
The atmospheric parameters were derived using the method of excitation and ionization  equilibrium of \ion{Fe}{I} and \ion{Fe}{II} lines and from those atmospheric parameters we calculated the abundances for other elements and found that none of the elements measured presents star-to-star variation greater than the uncertainties. We compared the derived abundances with those of other globular clusters and field stars available in the literature. We determined a mean [Fe/H] = $-$0.60 $\pm$ 0.03 for NGC\,6366 and found some similarity of this object with M\,71, another inner halo globular cluster. The Na-O anticorrelation extension is short and no star-to-star variation in Al is found. The presence of second generation stars is not evident in NGC\,6366.

\end{abstract}


\section{Introduction}
\label{sec:intro}

Globular clusters (GCs) were, for many years, viewed as the prototypes of simple stellar populations, that is, all their stars would have the same age and metallicity in a first approximation. Nevertheless, it was known for decades that GCs like 47\,Tucanae have abundance variations in some light elements, which could be explained by primordial origin or some unknown mixing mechanism~\citep[e.g.][]{Hesser80}. A paradigm shift began to occur when the Na-O and Mg-Al anticorrelations, already known for Red Giant Branch (RGB) stars~\citep[e.g.][]{Gratton04,Gratton12}, were also found in Sub-Giant Branch (SGB) and Main-Sequence (MS) stars that did not suffer any mixing process~\citep[e.g.][]{Gratton01}. Since stellar evolution models do not account for these observational results, the hypothesis of multiple stellar populations gained strength. Afterwards, multiple populations began to be identified in high precision photometric observations, first in $\omega$ Centauri~\citep{Bedin04}, which is the brightest Galactic GC, and new photometric techniques are being developed to study GCs multiple populations~\citep[see e.g.][]{Milone17}. As of today, the consensus is that some of the light elements (Li, CNO, Na, Al, Mg and F) studied in GCs present star-to-star abundance variations. There is no consensus yet about the nature of the polluters, with the main candidates being intermediate-mass Asymptotic Giant Branch (AGB) stars~\citep{Ventura01}, fast rotating massive stars~\citep{Decressin07}, massive interacting binaries~\citep{deMink09} and supermassive stars~\citep{Denissenkov14}. However,~\citet{Renzini15} argues that three of these four models violate the constraints imposed by the observations, and only the AGB scenario could explain the multiple populations, but only if its flaws -- like the difficulty in reproducing the observed Na-O anticorrelation -- can be fixed.

Despite being important for calibration of metallicity and abundance ratio studies based on low-resolution spectra and/or population synthesis of star clusters, few metal-rich Galactic GCs have been studied with high-resolution and high S/N ratio spectra to date. In this sense, NGC\,6366 (Galactic coordinates l = 18.4085~deg, b = +16.0356~deg) is an interesting object to be investigated. It is the fifth closest GC according to the catalogue of~\citet[][2010 edition]{Harris96}, with R$\sun$ = 3.5~kpc. Using isochrone fitting,~\citet{Campos13} found (m-M)$_V$ = 15.02 with A$_V \approx$ 2.11. Also, this GC still has no measurements based on high resolution spectra of chemical elements widely studied in the context of Galactic and GC chemical evolution, such as Al and Eu.

Previous photometric and spectroscopic studies have found a range of [Fe/H] values for NGC\,6366. The first determination was from~\citet{Pike76}, [M/H] = $-$0.50 $\pm$ 0.20, using the~\citet{Hartwick68} S parameter, derived from the colour-magnitude diagram (CMD) morphology. Later,~\citet{DaCosta95} found [Fe/H] = $-$0.65 $\pm$ 0.07 using fits in the \ion{Ca}{II} triplet in the near infrared, while a more recent analysis based on low resolution data from~\citet{Dias16} measured [Fe/H] = $-$0.61 $\pm$ 0.05, through spectrum fitting from two different spectral libraries. ~\citet{Johnson16} published the first high-resolution (R $\approx$ 38000) work on NGC\,6366. From atmospheric model measurements of red giants, they estimated [Fe/H] = $-$0.55 for this cluster. The mean value given in the~\citeauthor{Harris96} catalogue is [Fe/H] = $-$0.59.

Being metal-rich and near the outer bulge, NGC\,6366 may be chemically associated to this Galactic component. Usually, bulge GCs have [Fe/H] $> -$1.3~\citep{Bica16}. Also, they pass frequently in the high-density regions of the Galaxy, being subject to tidal stripping, which seems to be the case of NGC\,6366~\citep{Paust09}. According to~\citet{DaCosta89}, NGC\,6366 has halo kinematics, and is 'unambiguously' associated with that Galactic component, so it could be either a metal-rich GC from the inner halo or a bulge GC which suffered kinematic heating while evolving with the Galaxy.

~\citet{Campos13} calculated an age of 11 $\pm$ 1.15 Gyr for this GC, pointing out that the presence of multiple populations or CNO enhancement could affect the age estimation. However, the presence or absence of multiple stellar populations in NGC\,6366 is currently an open question. Based on a photometric index, (\emph{U}-\emph{B})-(\emph{B}-\emph{I}), which is sensitive to the abundances of helium and light elements,~\citet[][]{Monelli13} suggest a bimodality in the RGB of NGC\,6366.

The goal of this paper is to present a detailed chemical abundance analysis of bright giant stars in NGC\,6366 based on the highest resolution spectra obtained to date for this GC. We have derived chemical abundances for sixteen species (O, Na, Al, Si, Ca, Sc, Ti, V, Cr, Mn, Fe, Co, Ni, Y, Ba and Eu), including the first measurement of Al, that is crucial for the analysis of multiple populations. Our work provides additional observational constraints to the understanding of the chemical evolution of the Galaxy and of the GC chemical evolution itself.

In Section~\ref{sec:observ} we describe the observations and the data reduction process. Section~\ref{sec:analysis} presents the method of determination of radial velocities, atmospheric parameters and chemical abundances. Results and discussion of the dataset are presented in Section~\ref{sec:results} and, in Section~\ref{sec:conclusion}, we make final remarks.

\section{Observations and data reduction}
\label{sec:observ}

\begin{table*}
\caption{Program stars, coordinates, distance to the geometric centre of the GC and photometry used (see text for references).}
\label{tab:basic}
\begin{tabular}{ccccccccc}
\hline
ID & ID & RA & Dec & r$_{\mathrm{pos}}$ & \emph{V} & \emph{J} & \emph{H} & \emph{K$_s$} \\
(this work)   & \citep{Dias16} & (hh:mm:ss) & (dd:mm:ss) & (arcmin) & (mag) & (mag) & (mag) & (mag) \\
\hline
\multicolumn{8}{c}{UVES@VLT}
\\
\hline
01 & $\cdots$ & 17 27 58.32 & $-$05 04 29.35 & 3.53 & 15.731 & 12.517 & 11.837 & 11.680 \\
03 & $\cdots$ & 17 27 57.79 & $-$05 08 15.40 & 4.85 & 15.607 & 12.107 & 11.316 & 11.078 \\
05 & $\cdots$ & 17 27 54.84 & $-$05 01 54.80 & 3.91 & 15.807 & 12.472 & 11.760 & 11.597 \\
06 & $\cdots$ & 17 27 40.93 & $-$05 02 01.93 & 2.88 & 15.650 & 12.398 & 11.677 & 11.495 \\
07 & $\cdots$ & 17 27 34.30 & $-$05 01 09.66 & 4.40 & 15.858 & 12.468 & 11.777 & 11.555 \\
08 & $\cdots$ & 17 27 27.03 & $-$05 03 16.70 & 4.56 & 15.713 & 12.464 & 11.812 & 11.622 \\
\hline
\multicolumn{8}{c}{ESPaDOnS@CFHT}
\\
\hline
11 &   15     & 17 27 43.69 & $-$05 06 30.30 & 1.72 & 14.305 & 10.637 &  9.787 &  9.589 \\
18 &   05     & 17 27 44.48 & $-$05 02 38.10 & 2.16 & 14.697 & 10.985 & 10.157 &  9.946 \\
\hline
\end{tabular}
\end{table*}

\begin{figure}
	\includegraphics[width=\columnwidth]{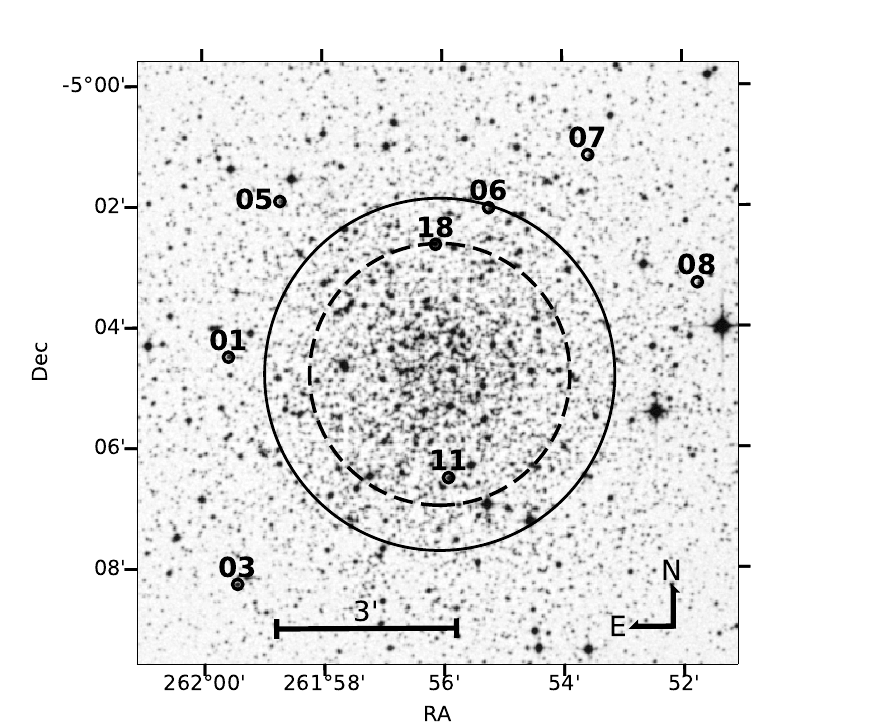}
    \caption{Finding chart for NGC\,6366 with the sample stars labeled. The dashed and solid circumferences delimit, respectively, the core radius and the half-light radius, taken from~\citet[][2010 edition]{Harris96}. Image from the Digitized Sky Survey.}
    \label{fig:fc}
\end{figure}

Our sample consists of eight bright giant stars (see Table~\ref{tab:basic} and Fig.~\ref{fig:fc} for details). The two brightest targets -- stars 11 and 18 -- were selected from~\citet{Pike76}, where their IDs are III-11 and I-18, respectively. Their spectra were acquired with ESPaDOnS, which is a high-dispersion echelle spectrograph and spectropolarimeter mounted at the 3.6-m Canada France Hawaii Telescope (CFHT). The observations were carried out on two nights of March and July 2015, in the star+sky mode. The exposure times were 3 x 2315\,s per star, reaching R $\sim$ 68,000. The observed wavelength ranges from ultraviolet (3699~\AA) to near-infra-red wavelengths (10483~\AA). ESPaDOnS data are reduced through the software Upena/Libre-Esprit pipeline, provided by the instrument team~\citep{Donati97}. Both stars 11 and 18 are also included in the analysis of~\citet{Dias16}, where their IDs are 15 and 05, respectively.

\begin{figure}
	\includegraphics[width=\columnwidth]{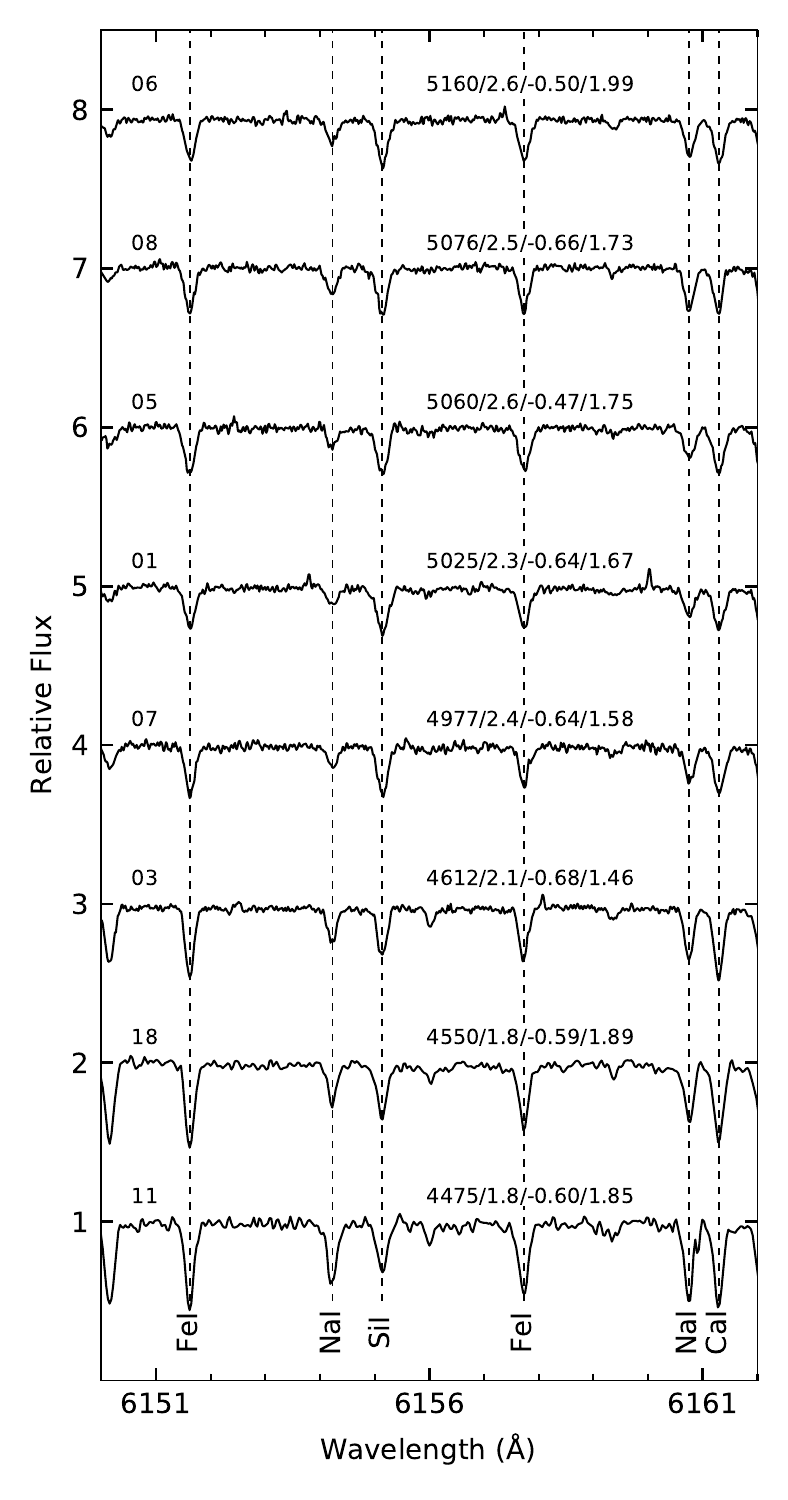}
    \caption{A portion of the spectra with some of the lines measured in this work identified. Above each spectrum are the IDs of the stars and the respective spectroscopic atmospheric parameters T$_{\mathrm{eff}}$ (K), log $g$ (dex), [Fe/H] (dex) and $\xi_{\mathrm{t}}$ (km s$^{-1}$). The spectra are ordered by decreasing T$_{\mathrm{eff}}$.}
    \label{fig:spec}
\end{figure}

The remaining targets were acquired from the ESO archive. They were observed with the Ultraviolet and Visual Echelle Spectrograph (UVES)~\citep{Dekker00} at the Very Large Telescope (VLT, four 8.2-m telescopes) for a study of metal-rich Galactic GCs~\citep{Feltzing09}. An 1 arcsec slit was used, yielding a resolution R $\sim$ 40,000. The exposure times are 2 $\times$ 3300\,s, except in the blue region (4727-5804~\AA) of stars 05 and 06, which have a single exposure each. For stars 03 and 07 only the red portion of the spectra, which covers wavelengths from 5818 to 6835~\AA, is available. All spectra were taken directly from the archive. We only needed to normalize the UVES spectra and to correct all of them for Doppler shift. These tasks were made with standard {\sc iraf} routines. After proceeding with the Doppler shift correction, spectra with multiple exposures were co-added to increase the signal-to-noise ratio (S/N) of individual stars in both subsamples. Typical values of S/N are 80 per pixel for the UVES@VLT subsample and 60 per pixel for the ESPaDOnS@CFHT stars, measured at $\lambda \approx$ 6150~\AA. We estimate S/N $\sim$ 25 at $\lambda \approx$ 5000~\AA\ in ESPaDOnS spectra -- see Section~\ref{sec:analysis} for a discussion about the implications of lower S/N. Fig.~\ref{fig:spec} shows a portion of all spectra.

\begin{figure}
	\includegraphics[width=\columnwidth]{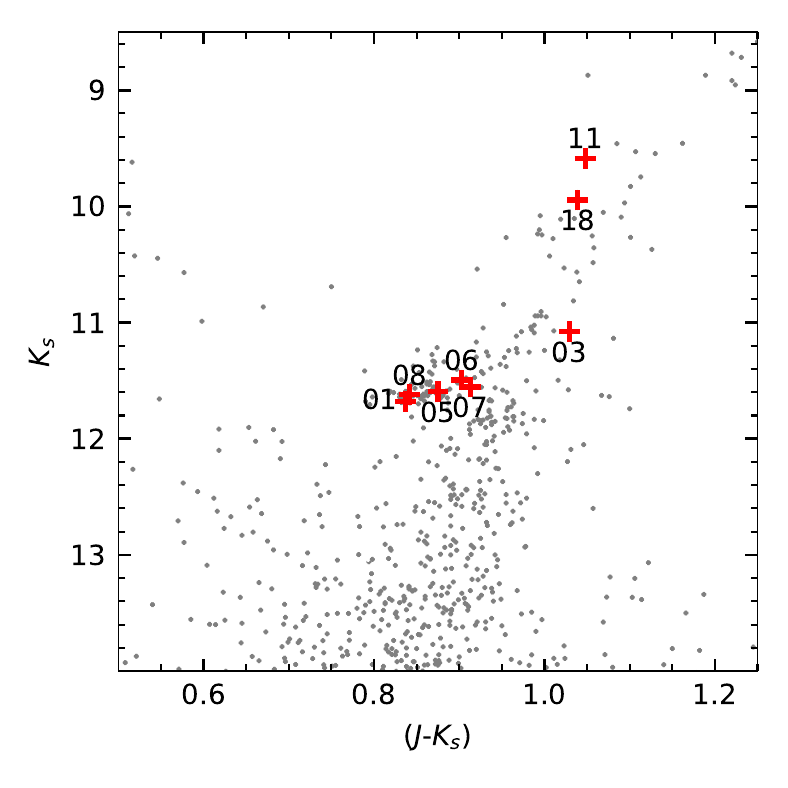}
    \caption{A \emph{Ks} vs. (\emph{J}-\emph{K$_{s}$}) CMD of NGC\,6366. \emph{Red crosses}: stars studied in this work. \emph{Grey dots}: 2MASS point sources within 10~arcmin from the centre of the GC.}
    \label{fig:cmd}
\end{figure}

The spatial distribution of the sample is shown in Fig.~\ref{fig:fc}, while in Fig.~\ref{fig:cmd} we show the location of the sample in a \emph{K$_s$} vs. (\emph{J}-\emph{K$_s$}) CMD. The photometry of \emph{J}, \emph{H} and \emph{K$_s$} bands is from the Two Micron All Sky Survey (2MASS) catalog~\citep{Skrutskie06}. In the \emph{V}-band, the photometric data for star 06 is from the  Naval Observatory Merged Astrometric Dataset (NOMAD/YB6) catalog~\citep{Zacharias04}. For stars 01 and 11, the \emph{V} photometry is from~\citet{Stetson00}. Otherwise, the photometric data in the \emph{V}-band is from~\citet{Sariya15}.

\section{Analysis}
\label{sec:analysis}

The heliocentric radial velocities were determined using the {\sc iraf} \emph{rvidlines} task, by measuring the positions of a large number of atomic absorption lines in each spectrum. In the UVES subsample, the adopted values are from the red part of each spectrum -- because two stars lack the blue part, as pointed out previously -- with 69 absorption features measured. In the ESPaDOnS subsample, 217 lines were measured along all spectra. For each object, the final value of radial velocity is the mean of the values obtained for each exposure. The uncertainties of the radial velocities of each star were propagated from the standard deviation of the mean of the measurements from each exposure.

The stellar atmospheric parameters -- effective temperature (T$_{\mathrm{eff}}$), surface gravity (log $g$), metallicity ([Fe/H]), and microturbulent velocity ($\xi_{\mathrm{t}}$) -- which are essential in a chemical abundance analysis, were obtained by photometric (for T$_{\mathrm{eff}}$ and log $g$) and spectroscopic methods (for all). The photometric T$_{\mathrm{eff}}$ were calculated using de-reddened (\emph{J}-\emph{K$_{s}$}) and (\emph{V}-\emph{$K_{s}$}) colours employing the calibrations of~\citet[][see \citet{Alonso01} for an erratum]{Alonso99}. The colour-temperature (\emph{V}-\emph{K$_{s}$}) calibration was chosen for first guesses because it has the lowest mean variation among all calibrations -- an error of 0.05 in magnitude implies an error of 0.7 percent in T$_{\mathrm{eff}}$. As the colour-temperature (\emph{J}-\emph{K}) calibration is independent of [Fe/H], its effective temperatures and surface gravity values were used as a control group. The 2MASS colours were transformed to the CIT system~\citep{Carpenter01} and then to the TCS system~\citep{Alonso98}. We adopted E(\emph{B}-\emph{V}) = 0.71 and, initially, [Fe/H] = $-$0.59~\citep[][2010 edition]{Harris96}. The relations E(\emph{V}-\emph{K})/E(\emph{B}-\emph{V}) = 2.744 and E(\emph{J}-\emph{K})/E(\emph{B}-\emph{V}) = 0.525~\citep{Rieke85} were used for extinction estimation. \citet{Campos13} built a detailed reddening map of NGC\,6366, but the map comprises the central region of the GC only, thus we have no detailed information about differential reddening in the outer regions where our sample stars are located, and a constant value of E(\emph{B}-\emph{V}) was chosen.

To estimate photometric log $g$ we used the relation:

\begin{equation}
\mathrm{log}(g_{\mathrm{*}}) = \mathrm{log}\left (\frac{\mu_{\mathrm{*}}}{\mu_{\sun}} \right) + 4 \mathrm{log}\left (\frac{T_{\mathrm{eff*}}}{T_{\mathrm{eff}\sun}} \right) + \frac{2}{5}(M_{\mathrm{bol},\mathrm{*}} - M_{\mathrm{bol},\sun}) + \mathrm{log}(g_{\sun}),
\end{equation}

\noindent where $\mu_{\mathrm{*}}$ is stellar mass and $M_{\mathrm{bol}}$ the bolometric magnitude. To perform the calculations, we assumed mass of 0.8 M$_{\odot}$ for all stars from inspection of an 11-Gyr PARSEC isochrone~\citep{Bressan12} with global metallicity Z=0.0039~\footnote{Z$_{\sun} =$ 0.0152.} and RGB mass loss coefficient $\eta_{\mathrm{Reimers}}$ = 0.2. The surface gravities of the Horizontal Branch (HB) stars change by 0.06~dex if we change $\eta_{\mathrm{Reimers}}$ from 0.2 to 0.4. The bolometric magnitude was estimated with the method described in~\citet[][Eq. (7)]{Torres10}. The values adopted for BC$_{V,\sun}$ and $M_{\mathrm{bol},\sun}$ were $-$0.08 and 4.76, respectively~\citep[][with $M_{\mathrm{bol},\sun}$ corrected as recommended by~\citeauthor{Torres10}]{Allen76}.

After the derivation of the [Fe/H] for each star using the spectroscopic method, described hereafter, we recalculated the (\emph{V}-\emph{K}) effective temperatures and surface gravities inserting the [Fe/H] values derived spectroscopically into the colour-T$_{\mathrm{eff}}$ calibration, star-by-star. As the (\emph{J}-\emph{K}) calibration is metallicity-independent, it was not necessary to recalculate its photometric parameters.

The spectroscopic stellar parameters for each object were determined by traditional methods. Equivalent widths (EW) of \ion{Fe}{I} and \ion{Fe}{II} lines were measured by hand using {\sc iraf} \emph{splot} task. Those features with EW > 160 m\AA\ were discarded, since stronger lines do not possess the gaussian profiles needed for our analysis. For weaker lines, lower S/N makes the measurement difficult. If we define the S/N of a normalized spectrum in a given region as the inverse of the standard deviation (1-$\sigma$) of the continuum noise, and assume that the minimum depth for a line to be distinguishable from continuum noise is 3-$\sigma$, the minimum EW for a gaussian absorption feature be considered in our analysis is $\sim$ 19~m\AA\ at S/N = 25 ($\lambda \approx$ 5000~\AA\ in ESPaDOnS spectra, as discussed in Section~\ref{sec:observ}) for a typical FWHM of 150~m\AA. No lines with EW $<$ 19~m\AA\ were considered for the stars observed with ESPaDOnS, and no lines with EW $<$ 40~m\AA\ were taken into account in the bluest ($\lambda <$ 5000~\AA) regions of their spectra.

\begin{figure}
	\includegraphics[width=\columnwidth]{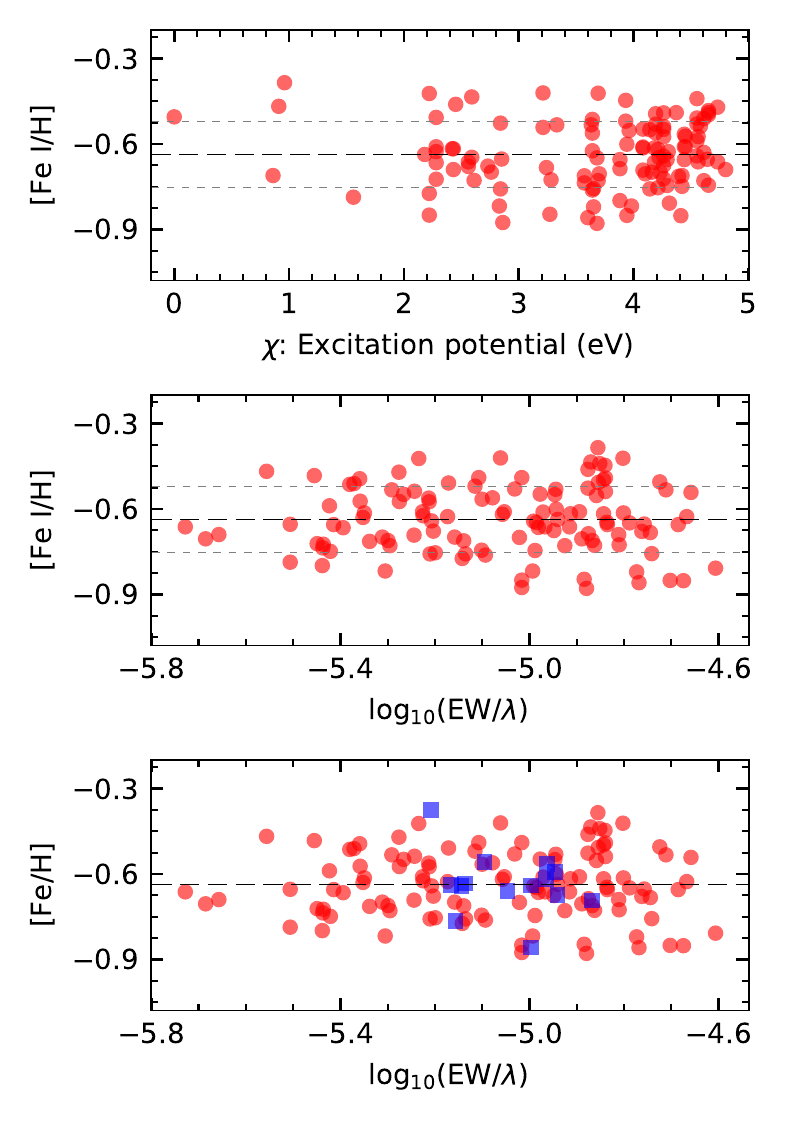}
    \caption{
    Diagrams showing the end of the iterative process used to find the spectroscopic atmospheric parameters for star 01.
    \emph{Top}: Excitation equilibrium ($\chi$ vs. abundance) diagram, employed to find the spectroscopic effective temperature when the slope of the least-squares fit is zero.
    \emph{Centre}: Reduced width (EW/$\lambda$) vs. abundance diagram. Used to find the microturbulence velocity also when the slope of the least-squares fit is zero.
    \emph{Bottom}: Ionization equilibrium (\ion{Fe}{I} vs. \ion{Fe}{II}) diagram. Employed to find the spectroscopic surface gravity when abundances of both \ion{Fe}{I} and \ion{Fe}{II} species are equal, where red circles and blue squares represent \ion{Fe}{I} and \ion{Fe}{II} absorption features, respectively. Black dashed lines are the least-squares fit -- when their slope is zero their y-value equals \ion{Fe}{I} abundance. Grey dotted lines are 1-$\sigma$ limits. In the bottom plot, the black dotted line indicates the least squares fit to both \ion{Fe}{I} and \ion{Fe}{II} points.}
    \label{fig:eqexc}
\end{figure}

For each star, a LTE 1D plane-parallel atmospheric model was built with the ATLAS9 code~\citep{Castelli97}, assuming a solar $\alpha$ mixture, and used as input in the \emph{abfind} driver of {\sc moog}~\citep{Sneden73}. The expected difference in abundances between plane-parallel and spherical models is negligible ($\sim$~0.02 dex; see, e.g.,~\cite{AlvesBrito10}). The effective temperature (T$_{\mathrm{eff}}$) was found by removing any trends in plots of \ion{Fe}{I} abundance versus line excitation potential ($\chi$) -- excitation equilibrium. The same approach was used in plots of \ion{Fe}{I} abundance versus reduced equivalent width (log (EW/$\lambda$)) to find the microturbulent velocity $\xi_{\mathrm{t}}$. By forcing the abundances of \ion{Fe}{I} and \ion{Fe}{II} to be equal we obtained the surface gravity in terms of the ionization equilibrium. Finally, the metallicity [Fe/H] was determined by equalizing the value used in the model and the value given by {\sc moog} in the output. The adopted stellar parameters were found when all the conditions were satisfied simultaneously, after an iterative process. Examples of excitation equilibrium diagrams are given in Fig.~\ref{fig:eqexc}. The \ion{Fe}{I} abundances are subject to NLTE effects which could affect the values of the atmospheric parameters, particularly the surface gravity. However, the use of the LTE approximation is justified as these effects are negligible for cold giants in the considered metallicity regime ([Fe/H] $\sim$ -0.60), and NLTE corrections would be small if compared to the adopted uncertainties~\citep[see Figs. 2, 4, 6 and 7 of][and references therein]{Lind12}.

\begin{figure}
	\includegraphics[width=\columnwidth]{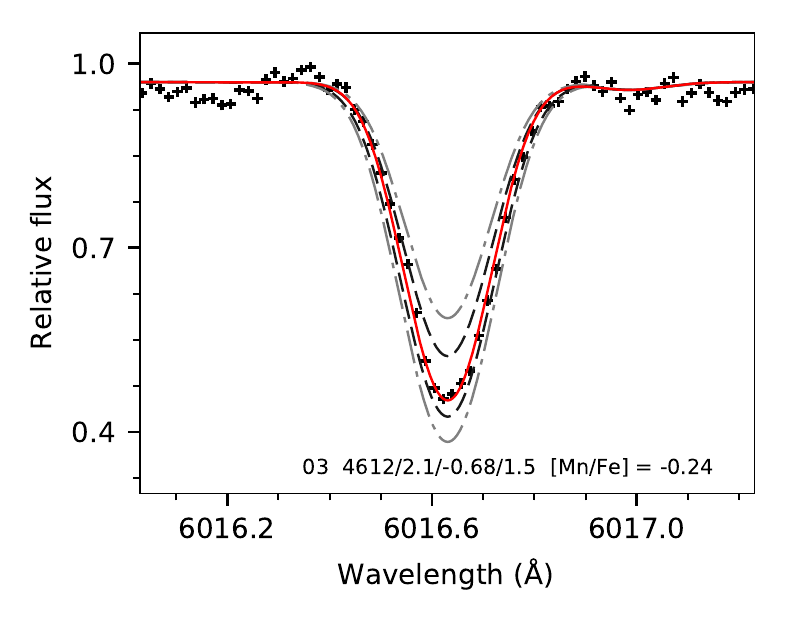}
    \caption{Example of spectral synthesis of the 6016-\AA\ line of Mn. The points are the observed spectra. The solid red line is the best-fitting, while the grey lines represent the upper and lower uncertainties in [Mn/Fe] (dashed dark grey for 1-$\sigma$, dash-dot light grey for 2-$\sigma$). Star ID, the four atmospheric parameters and the mean abundance of all lines measured for the star are labelled in the figure.}
    \label{fig:synth_Mn}
\end{figure}

To obtain the abundance of elements with atomic number Z $\leq$ 28 we adopted EW measurements used as input in the \emph{abfind} driver of {\sc moog}, except for Mn. For heavier elements and Mn, the spectral synthesis method was used by employing the \emph{synth} driver of {\sc moog}. The spectral synthesis method consists of comparing absorption lines of synthetic and observed spectra, where the former are built from the chosen atmospheric model for each star. The chemical abundance of a given element X is found varying [X/Fe] in the input to make the synthetic and observed line profiles match. Fig.~\ref{fig:synth_Mn} shows a visualization of the method (see also Figs.~\ref{fig:synth_Y},~\ref{fig:synth_Ba} and~\ref{fig:synth_Eu}).

The adopted line list for all elements is the same as given in~\citet{Yong14}, except for Mn, where we adopted the hyperfine structure derived by~\citet{Barbuy13} for the Mn 6000-\AA\ triplet. For Eu we used the same hyperfine structure of~\citet[][sent by private communication]{Marino11}. Table~\ref{tab:ew} presents the atomic data of log \emph{gf} and $\chi$ and EW measurements for individual lines of the elements up to Ni. The chemical abundances of the lines measured by spectral synthesis are in Table~\ref{tab:ncapture}.

\begin{table*}
\caption{Atomic line list: wavelength, ionization state (integer part: atomic number; decimal part: .0 for neutral and .1 for single ionized species), atomic parameters $\chi$ and log \emph{gf} and equivalent widths for each star. This table is published in its entirety in the electronic edition of the paper. See
text for sources of the atomic parameters.}
\label{tab:ew}
\begin{tabular}{cccrrrrrrrrr}
\hline
Wavelength & Ion & $\chi$ & log \emph{gf} & 01     & 03     & 05     & 06     & 07     & 08     & 11     & 18     \\
(\AA)      &     &  (eV)  &               & (m\AA) & (m\AA) & (m\AA) & (m\AA) & (m\AA) & (m\AA) & (m\AA) & (m\AA) \\
\hline
6300.31 & 8.0  & 0.00 & $-$9.75  & 33.4 & 39.5 & 36.5 & 34.0 & 31.5 & 29.9 & 49.1 & 52.0 \\
6363.78 & 8.0  & 0.02 & $-$10.25 & 12.0 & 17.0 & 19.2 & 9.3 & 13.9 & 11.8 & $\cdots$ & 26.1 \\
5682.65 & 11.0 & 2.10 & $-$0.67  & 79.8 & $\cdots$ & 96.2 & 106.1 & $\cdots$ & 106.6 & 148.7 & 131.5 \\
5688.22 & 11.0 & 2.10 & $-$0.37  & 101.4 & $\cdots$ & 112.8 & 116.6 & $\cdots$ & 118.7 & 157.6 & 139.3 \\
6154.23 & 11.0 & 2.10 & $-$1.57  & 32.1 & 46.9 & 30.9 & 38.9 & 36.3 & 45.3 & 88.4 & 52.4 \\
6160.75 & 11.0 & 2.10 & $-$1.26  & 37.9 & 63.4 & 50.6 & 57.3 & 50.1 & 57.0 & 103.0 & 78.1 \\
$\cdots$ & $\cdots$  & $\cdots$ &  $\cdots$  & $\cdots$ & $\cdots$ & $\cdots$ & $\cdots$ & $\cdots$ & $\cdots$ & $\cdots$ & $\cdots$ \\
\hline
\end{tabular}
\end{table*}

\begin{table*}
\caption{Line-by-line abundances obtained by the spectral synthesis method.}
\begin{tabular}{lcrrrrrrrr}
\hline
[X/Fe] & wavelength (\AA) & 01 & 03 & 05 & 06 & 07 & 08 & 11 & 18 \\
\hline
$[$Mn/Fe$]$ & 6013.51 & $-$0.35 & $-$0.26 & $-$0.40 & $-$0.33 & $-$0.17 & $-$0.25 & $-$0.24 & $-$0.30  \\
$[$Mn/Fe$]$ & 6016.64 & $-$0.31 & $-$0.18 & $-$0.36 & $-$0.37 & $-$0.18 & $-$0.31 & $-$0.40 & $-$0.33  \\
$[$Mn/Fe$]$ & 6021.80 & $-$0.33 & $-$0.28 & $-$0.42 & $-$0.44 & $-$0.25 & $-$0.27 & $-$0.35 & $-$0.21  \\
$[$Y/Fe$]$  & 5087.42 & $-$0.40 & $\cdots$ & $-$0.30 & $-$0.30 & $\cdots$ & $-$0.10 & $-$0.05  & $-$0.10  \\
$[$Y/Fe$]$  & 5200.42 & $-$0.25 & $\cdots$ & $-$0.12 & $-$0.20 & $\cdots$ & $-$0.15 & $-$0.15  & 0.01     \\
$[$Ba/Fe$]$ & 5853.69 & $-$0.22 & $-$0.04  & $-$0.15 & $-$0.20 & $-$0.06  & $-$0.01 & 0.00     & $-$0.18  \\
$[$Eu/Fe$]$ & 6645.06 & 0.15 & 0.33 & 0.16 & 0.38 & 0.23 & 0.15 & 0.30 & 0.23 \\
\hline
\end{tabular}
\label{tab:ncapture}
\end{table*}

\subsection{Uncertainties}
\label{sec:unc}

For each star, after converging the values of the atmospheric parameters to the adopted figures, the abundances of Fe show a dispersion around the mean (see Fig.~\ref{fig:eqexc}). To estimate the uncertainties of each atmospheric parameter, we varied one atmospheric parameter at a time until the abundance of \ion{Fe}{I} returned in the {\sc moog} output changed by a standard deviation. The atmospheric parameters uncertainties adopted in this work are typical values found in these estimates. The values are $\Delta$T$_{\mathrm{eff}} \approx \pm$ 150~K, $\Delta$log $g$ $\approx \pm$ 0.3~dex, $\Delta$[Fe/H] $\approx \pm$ 0.1~dex and $\Delta \xi_{\mathrm{t}} \approx \pm$ 0.2~km~s$^{-1}$. To estimate the uncertainties in chemical abundances, we created atmospheric models varying only one parameter by its uncertainty value per new model created. The sensitivities of the abundance values due to each atmospheric parameter variation were treated as the uncertainties for each element related to that atmospheric parameter variation and were propagated by quadrature. Then, the value was propagated again by quadrature with the standard deviation of line-by-line abundances given by {\sc moog} to get the total uncertainties for elements measured from EW (see Table~\ref{tab:unc}).

The sensitivities of the abundances to variations of the stellar atmospheric parameters of two representative stars -- one from the HB and another one from the cold giants -- are shown in Table~\ref{tab:sensitiv}. The uncertainties for elements measured by spectral synthesis are from these two representative stars: the uncertainties for Mn, Y, Ba and Eu of the UVES subsample were taken from the sensitivities of star 01, while the uncertainties for  these elements of the ESPaDOnS subsample were taken from sensitivities of star 18. We also defined the ratio $\rho = \sigma_{\mathrm{cl}}/\langle \sigma_{\mathrm{*}} \rangle$ where $\sigma_{\mathrm{cl}}$ and $\langle \sigma_{\mathrm{*}} \rangle$ are, respectively, the standard deviation of the star-to-star variation of a certain element in the GC and the mean uncertainty of this element.

\begin{table}
\caption{Sensitivities to 1-$\sigma$ variation of the atmospheric parameters in the abundance ratios in stars 01 and 18. See text for uncertainties in the atmospheric parameters.}
\label{tab:sensitiv}
\begin{tabular}{lrrrr}
\hline
Element & $\Delta$T$_{\mathrm{eff}}$ & $\Delta$log $g$ & $\Delta$[Fe/H] & $\Delta \xi_{t}$ \\
 & $\pm$150 &  $\pm$0.3  &  $\pm$0.1  &     $\pm$0.2       \\
 & (K) & (dex) & (dex) & (km s$^{-1}$) \\
\hline
\multicolumn{5}{c}{01}
\\
\hline
\ion{Fe}{I} &   0.15&$<$0.01&$<$0.01&   0.05 \\
\ion{Fe}{II}&   0.06&   0.14&   0.02&   0.05 \\
\ion{O}{I}  &   0.04&   0.13&   0.03&$<$0.01 \\
\ion{Na}{I} &   0.11&   0.02&$<$0.01&   0.03 \\
\ion{Mg}{I} &   0.11&   0.01&$<$0.01&   0.04 \\
\ion{Al}{I} &   0.10&$<$0.01&$<$0.01&$<$0.01 \\
\ion{Si}{I} &   0.05&   0.03&$<$0.01&   0.02 \\
\ion{Ca}{I} &   0.14&   0.02&$<$0.01&   0.06 \\
\ion{Sc}{II}&$<$0.01&   0.12&   0.03&   0.06 \\
\ion{Ti}{I} &   0.21&$<$0.01&$<$0.01&   0.03 \\
\ion{Ti}{II}&$<$0.01&   0.12&   0.02&   0.09 \\
\ion{V}{I}  &   0.24&$<$0.01&$<$0.01&   0.02 \\
\ion{Cr}{I} &   0.18&   0.01&$<$0.01&   0.06 \\
\ion{Mn}{I} &   0.17&$<$0.01&   0.10&   0.01 \\
\ion{Co}{I} &   0.15&   0.01&$<$0.01&   0.02 \\
\ion{Ni}{I} &   0.15&   0.01&$<$0.01&   0.04 \\
\ion{Y}{II} &   0.06&   0.10&   0.11&   0.08 \\
\ion{Ba}{II}&   0.09&   0.15&   0.03&   0.08 \\
\ion{Eu}{II}&   0.05&   0.23&$<$0.01&   0.05 \\
\hline
\multicolumn{5}{c}{18}
\\
\hline
\ion{Fe}{I} &0.11&   0.04&$<$0.01&   0.06 \\
\ion{Fe}{II}&0.15&   0.17&   0.03&   0.04 \\
\ion{O}{I}  &0.02&   0.13&   0.04&$<$0.01 \\
\ion{Na}{I} &0.14&   0.02&$<$0.01&   0.05 \\
\ion{Mg}{I} &0.09&$<$0.01&$<$0.01&   0.06 \\
\ion{Al}{I} &0.11&$<$0.01&$<$0.01&   0.02 \\
\ion{Si}{I} &0.05&   0.07&   0.01&   0.03 \\
\ion{Ca}{I} &0.16&   0.01&$<$0.01&   0.09 \\
\ion{Sc}{II}&0.02&   0.13&   0.03&   0.08 \\
\ion{Ti}{I} &0.23&$<$0.01&$<$0.01&   0.06 \\
\ion{Ti}{II}&0.03&   0.13&   0.03&   0.09 \\
\ion{V}{I}  &0.28&   0.02&$<$0.01&   0.09 \\
\ion{Cr}{I} &0.16&$<$0.01&$<$0.01&   0.06 \\
\ion{Mn}{I} &0.18&   0.03&   0.09&   0.04 \\
\ion{Co}{I} &0.10&   0.05&   0.02&   0.04 \\
\ion{Ni}{I} &0.08&   0.06&   0.02&   0.07 \\
\ion{Y}{II} &0.02&   0.10&   0.08&   0.12 \\
\ion{Ba}{II}&0.04&   0.11&   0.05&   0.12 \\
\ion{Eu}{II}&0.03&   0.15&   0.05&   0.02 \\
\hline
\end{tabular}
\end{table}

\section{Results and discussion}
\label{sec:results}

In this section we present and discuss our results on kinematics and chemical abundances. The $\alpha$ elements, Fe-peak, light odd-Z elements, and those synthesised by neutron capture. The final abundance pattern will be compared with those displayed by stars in different environments, field and other GCs.

\subsection{Kinematics}
\label{sec:kin}

The mean heliocentric radial velocity (RV) for our sample is $-$121.2 $\pm$ 0.7~km~s$^{-1}$ ($\sigma$ = 2.0~km~s$^{-1}$). Star-by-star heliocentric radial velocities are outlined in Table~\ref{tab:rv}. The value from~\citet[][2010 edition]{Harris96} catalogue is $-$122.2~km~s$^{-1}$, with a central velocity dispersion of 1.3~km~s$^{-1}$. This result is also consistent with the value of $-$123.2 $\pm$ 1.0~km~s$^{-1}$ measured by~\citet{DaCosta89}, where four giants were analysed, whose results was used by the authors to conclude that NGC\,6366 \emph{has unambiguously halo kinematics}. The authors used the (V$_s$, $\Psi$) plane model, where V$_s$ is the radial velocity with respect to a stationary observer at the Sun and $\Psi$ is the angle between the line of sight to the cluster and the direction of Galactic rotation at the cluster ~\citep[see Fig. 1 from][where the model is explained]{Frenk80}. Using low-resolution spectra of 14 giants, \citet{Dias16} found a mean RV of $-$137 $\pm$ 54~km~s$^{-1}$. In their analysis, there are two objects in common with this work (stars 11 and 18) and the RVs for these individual objects agree in $<$~1~km~s$^{-1}$. Using the cross-correlation method,~\citet{Johnson16} found a mean heliocentric RV of $-$122.3~km~s$^{-1}$ ($\sigma$ = 1.5~km~s$^{-1}$) for NGC\,6366. The proper motion analysis of~\citet{Sariya15} gave zero membership probability for star 05 and membership probability $>$ 0.95 for stars 03, 07, 08 and 11. Stars 01, 06 and 11 were not present in their study. Based on our results for radial velocities, we assume cluster membership for all the eight stars, and the status of star 05 will be further discussed in the chemical analysis.

\begin{table}
\caption{Kinematic results.}
\label{tab:rv}
\begin{tabular}{cccr}
\hline
ID & RV & $\overline{\sigma}$ RV & \# of lines \\
     & (km s$^{-1}$) & (km s$^{-1}$) & measured \\
\hline
01 & $-$121.46 & 0.07 & 69 \\
03 & $-$119.23 & 0.05 & 69 \\
05 & $-$119.60 & 0.08 & 69 \\
06 & $-$119.06 & 0.07 & 69 \\
07 & $-$120.83 & 0.08 & 69 \\
08 & $-$125.17 & 0.06 & 69 \\
11 & $-$122.06 & 0.06 & 217 \\
18 & $-$121.80 & 0.05 & 217 \\
\hline
\end{tabular}
\end{table}

\subsection{Atmospheric parameters}

\begin{table*}
\caption{Atmospheric parameters derived by photometric and spectroscopic methods.}
\label{tab:atmpar}
\begin{tabular}{ccccccccc}
\hline
\multicolumn{1}{c}{} & 
\multicolumn{4}{c}{Spectroscopy} & 
\multicolumn{4}{c}{Photometry}
\\
ID & T$_{\mathrm{eff}}$ & log $g$ & [Fe/H] & $\xi_{\mathrm{t}}$ & T$_{\mathrm{eff},(V-K)}$ & log $g$$_{(V-K)}$ & T$_{\mathrm{eff},(J-K)}$ & log $g$$_{(J-K)}$ \\
 & (K) & (dex) & (dex) & (km s$^{-1}$) & (K) & (dex) & (K) & (dex) \\
\hline
01 & 5025 & 2.3 &$-$0.64 & 1.7 & 4985 & 2.4 & 5336 & 2.5 \\
03 & 4612 & 2.1 &$-$0.68 & 1.5 & 4516 & 2.0 & 4650 & 2.1 \\
05 & 5060 & 2.6 &$-$0.47 & 1.7 & 4817 & 2.3 & 5180 & 2.5 \\
06 & 5160 & 2.6 &$-$0.49 & 2.0 & 4874 & 2.3 & 5073 & 2.4 \\
07 & 4977 & 2.4 &$-$0.64 & 1.6 & 4719 & 2.3 & 5036 & 2.4 \\
08 & 5076 & 2.5 &$-$0.66 & 1.7 & 4939 & 2.3 & 5315 & 2.5 \\
11 & 4475 & 1.8 &$-$0.60 & 1.9 & 4373 & 1.4 & 4594 & 1.6 \\
18 & 4550 & 1.8 &$-$0.59 & 1.9 & 4348 & 1.5 & 4620 & 1.7 \\
\hline
\end{tabular}
\end{table*}

\begin{figure}
	\includegraphics[width=\columnwidth]{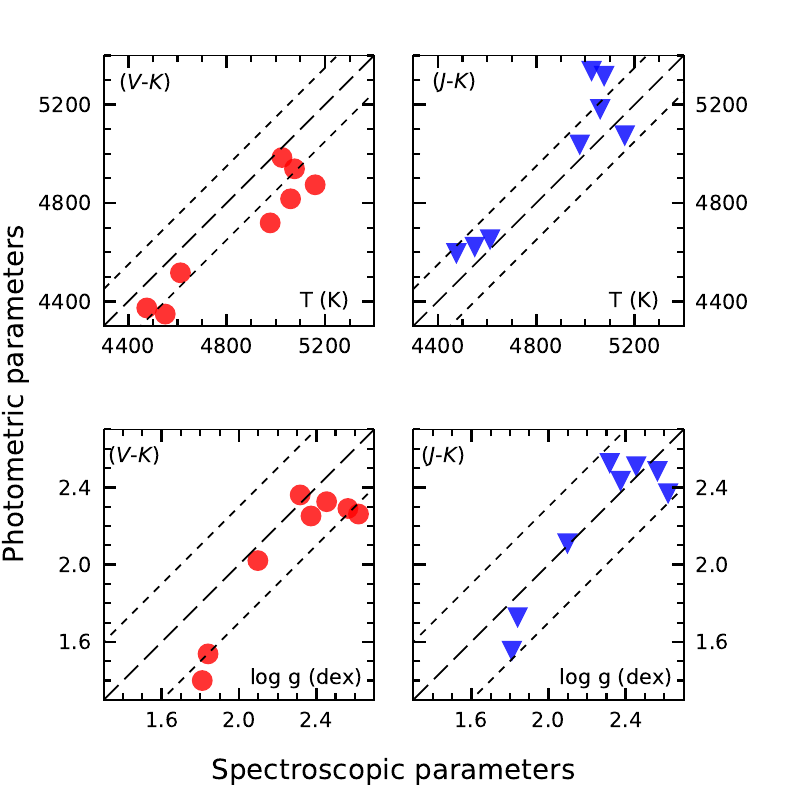}
    \caption{Comparison between atmospheric parameters obtained by spectroscopic (x-axis) and photometric (y-axis) methods. \emph{Red circles}: photometric parameters derived from the (\emph{V}-\emph{K}) calibration. \emph{Blue triangles}: photometric parameters derived from the (\emph{J}-\emph{K}) calibration. \emph{Dashed lines}: identity function. \emph{Dotted lines}: limits of the spectroscopic uncertainties adopted in this work (see Section~\ref{sec:observ}).}
    \label{fig:atmpar}
\end{figure}

In Fig.~\ref{fig:atmpar} and Table~\ref{tab:atmpar} the photometric and spectroscopic effective temperatures are compared. The difference between the mean spectroscopic T$_{\mathrm{eff}}$ and the mean T$_{\mathrm{eff},(V-K)}$ is 170~K ($\sigma =$ 89~K). For the mean T$_{\mathrm{eff},(J-K)}$ the difference is $-$108~K ($\sigma =$ 122~K). We recall that our adopted uncertainty is 150~K (see Section~\ref{sec:unc}). If spectroscopic and photometric measurements for individual stars are compared, the differences in T$_{\mathrm{eff}}$ can be as big as $\sim$300 K. As we adopted a fixed E(\emph{B}-\emph{V}) value, the main reason for the discrepancy can be differential reddening. Both \citet{Campos13} and \citet{Alonso97} found that reddening increases from south to north in NGC\,6366. If the two southernmost stars (03 and 11) are excluded, there is a trend between declination and the spectroscopic-photometric difference in both (\emph{V}-\emph{K}) and (\emph{J}-\emph{K}) calibrations. The spectroscopic-photometric difference increases from south to north, i.e., the photometric temperatures are ‘colder’ (redder) than they should be in the northernmost stars -- if the spectroscopic temperatures are taken as ‘true’ T$_{\mathrm{eff}}$. The $\Delta$E(\emph{V}-\emph{K}) should be $\sim$ 0.5~mag -- $\Delta$E(\emph{B}-\emph{V}) $\sim$ 0.18~mag using the relations given in Section~\ref{sec:analysis} -- to account for this trend alone in the (\emph{V}-\emph{K}) calibration, because an error of 0.01~mag corresponds to a 5~K variation in the photometric T$_{\mathrm{eff}}$~\citep{Alonso99}, and the spectroscopic-photometric difference varies from 35~K (star 01) to 286~K (star 06). Concerning the (\emph{J}-\emph{K}) calibration, if we assume the 23~K variation per 0.01~mag given by \citeauthor{Alonso99} for stars with (\emph{J}-\emph{K})~$>$~0.5, the $\Delta$E(\emph{J}-\emph{K}) should be 0.17~mag, or $\sim$ 0.3~mag in $\Delta$E(\emph{B}-\emph{V}), to account for the $\sim$ 400~K range in the spectroscopic-photometric difference in T$_{\mathrm{eff}}$ -- the difference is $-$311~K for star 01 and 87~K in star 06. However, the HB stars have (\emph{J}-\emph{K}) $\sim$ 0.5, and, for those bluer than this threshold, the variation in T$_{\mathrm{eff}}$ per 0.01~mag given by \citeauthor{Alonso99} is 69~K, resulting in $\Delta$E(\emph{B}-\emph{V}) $\sim$ 0.11~mag. A detailed reddening map of the entire area of the GC would be useful to clarify this issue, since the map from \citet{Alonso97} is a rough estimate and the map from \citet{Campos13}, the most detailed to date, covers only a small central field. The region mapped in \citeauthor{Campos13} shows a $\Delta$E(\emph{B}-\emph{V}) of 0.12~mag, indicating that our hypothesis of differential reddening as the main reason for the discrepancy in photometric and spectroscopic temperatures is feasible. Besides the differential reddening, there are the ‘shifts’ represented by the differences between mean photometric and mean spectroscopic temperatures, mentioned earlier in this paragraph, but those values -- comparable to the adopted uncertainties -- may be effect of using an unique value of E(\emph{B}-\emph{V}) for all stars, if our hypothesis of differential reddening influence is correct.

The log $g$ values are typical of giant stars~\citep[see, for example, Table 5 of][]{AlvesBrito10}. The difference of $\sim$ 0.9~dex between HB stars and the two most luminous ones -- those of the ESPaDOnS subsample -- is expected. Star 03 has an intermediary value of log $g$ = 2.1.

The microturbulent velocity is defined as a non-thermal component of the gas velocity on a smaller scale than the radial component of the line forming region~\citep{Cantiello09}, being responsible for a portion of line broadening. $\xi_{\mathrm{t}}$ values were derived only spectroscopically. The first guess value was arbitrarily set as 1.0~km~s$^{-1}$, and the final values found are typical for giants~\citep[see also][Table 5]{AlvesBrito10}.

As for metallicities, we found a mean [Fe/H] = $-$0.60 for NGC\,6366, with median [Fe/H] = $-$0.62. The mean metallicity is consistent with that found by~\citet{Johnson16} as well as studies based upon low-resolution data and photometry, mentioned in Section~\ref{sec:intro}. The two subsamples analysed in this work do not present different trends in metallicity. The mean uncertainty $\langle \sigma_{\mathrm{*}} \rangle$ found for [Fe/H] is 0.20~dex, while the standard deviation of star-to-star variation $\sigma_{\mathrm{cl}}$ is 0.08~dex. All [Fe/H] values are inside 1-$\langle \sigma_{\mathrm{*}} \rangle$ from the mean [Fe/H]. No metallicity variation was detected beyond uncertainties.

\begin{table*}
\caption{Chemical abundances (columns 2 to 9), in~dex -- where the solar photospheric abundances are from~\citet{Asplund09} -- the mean abundance of each element (column 10), its mean uncertainties (column 11), standard deviations of star-to-star variation (column 12) and the ratio $\rho = \sigma_{\mathrm{cl}}/\langle \sigma_{\mathrm{*}} \rangle$ in the last column.}
\label{tab:abund}
\begin{tabular}{lrrrrrrrrrrrrrr}
\hline
[X/Fe] & 01 & 03 & 05 & 06 & 07 & 08 & 11 & 18 & Mean &$\langle \sigma_{\mathrm{*}} \rangle$ & $\sigma_{\mathrm{cl}}$ & $\rho$ \\
\hline
\ion{O}{I}	& 0.46	& 0.47	 & 0.61	 & 0.47	 & 0.50	  & 0.52  & 0.35  & 0.47  & 0.48  &0.15 & 0.07	& 0.47 \\ 
\ion{Na}{I}	& 0.12	& 0.20	 & 0.10	 & 0.27	 & 0.23	  & 0.48  & 0.51  & 0.22  & 0.27  &0.15 & 0.15	& 1.00 \\ 
\ion{Mg}{I}	& 0.28	& 0.16	 & 0.23	 & 0.16	 & 0.18	  & 0.31  & 0.26  & 0.33  & 0.24  &0.15 & 0.07	& 0.47 \\ 
\ion{Al}{I}	& 0.31	& 0.44	 & 0.34	 & 0.32	 & 0.40	  & 0.39  & 0.29  & 0.41  & 0.36  &0.13 & 0.05	& 0.38 \\ 
\ion{Si}{I}	& 0.23	& 0.34	 & 0.20	 & 0.20	 & 0.32	  & 0.38  & 0.38  & 0.28  & 0.29  &0.12 & 0.08	& 0.67 \\ 
\ion{Ca}{I}	&$-$0.02& 0.23	 & 0.01	 & 0.03	 & 0.29	  & 0.14  & 0.07  & 0.08  & 0.10  &0.19 & 0.11	& 0.58 \\ 
\ion{Sc}{II}&$-$0.14& 0.12	 & 0.10	 &$-$0.02& 0.24	  & 0.10  & 0.08  & 0.11  & 0.08  &0.17 & 0.11	& 0.65 \\ 
\ion{Ti}{I}	& 0.19	& 0.33	 & 0.24	 & 0.33	 & 0.33	  & 0.28  & 0.23  & 0.25  & 0.27  &0.24 & 0.05	& 0.21 \\ 
\ion{Ti}{II}& 0.28	& 0.38	 & 0.24	 & 0.34	 & 0.33	  & 0.48  & 0.37  & 0.28  & 0.34  &0.18 & 0.08	& 0.44 \\ 
\ion{V}{I}	& 0.01	& 0.14	 & 0.05	 & 0.07	 & 0.08	  & 0.15  & 0.27  & 0.24  & 0.13  &0.28 & 0.09	& 0.32 \\ 
\ion{Cr}{I}	&$-$0.05&$\cdots$&$-$0.09&$-$0.03&$\cdots$&$-$0.07& 0.02  & 0.01  &$-$0.04&0.16 & 0.04	& 0.25 \\ 
\ion{Mn}{I}	&$-$0.33&$-$0.24 &$-$0.39&$-$0.38&$-$0.20 &$-$0.28&$-$0.33&$-$0.28&$-$0.30&0.20 & 0.07	& 0.35 \\ 
\ion{Co}{I}	& 0.05	& 0.07	 &$-$0.04& 0.02	 & 0.13	  & 0.07  & 0.18  & 0.16  & 0.08  &0.16 & 0.07	& 0.44 \\ 
\ion{Ni}{I}	& 0.01	& 0.06	 &$-$0.03& 0.03	 & 0.09	  & 0.02  & 0.05  & 0.02  & 0.03  &0.19 & 0.03	& 0.16 \\ 
\ion{Y}{II}	&$-$0.33	&$\cdots$&$-$0.21 &$-$0.25 &$\cdots$&$-$0.13 &$-$0.10 & $-$0.05 &$-$0.18&0.18 & 0.10	& 0.56 \\ 
\ion{Ba}{II}&$-$0.22&$-$0.04 &$-$0.15&$-$0.20&$-$0.06 &$-$0.01& 0.00  &$-$0.18&$-$0.11&0.18 & 0.09	& 0.50 \\ 
\ion{Eu}{II}& 0.15  & 0.33   & 0.16  & 0.38  & 0.23   & 0.15  & 0.30  & 0.23  & 0.24  &0.20 & 0.09	& 0.45 \\ 
\hline
\end{tabular}
\end{table*}

\subsection{Alpha elements}
\label{sec:alpha}

As discussed in~\citet{McWilliam16}, the $\alpha$ elements analysed in this work (O, Mg, Si, Ca and Ti) are grouped under the '$\alpha$' label because it was thought that they were produced by successive addition of $\alpha$ particles. However, more recent nucleosynthesis models for massive stars point to different channels of production for these elements, with only O having the $\alpha$ process as the main channel, while C and Ne hydrostatic burning dominate for Mg production. Explosive O burning is responsible for most Si and Ca abundances, while Ti is made in a variety of sites -- with the most abundant isotope, \ce{^{48}Ti} being produced by (explosive) incomplete Si burning and, to a lesser extent, type Ia Supernovae (SNe)~\citep[e.g.][]{WW95,Woosley02,Nomoto13}. Since these elements are released to the interstellar medium by massive stars, the [$\alpha$/Fe] ratio is a tracer of the IMF of the primordial gas that formed a stellar population~\citep[see e.g.][]{McWilliam97}.

The mean [O/Fe] for NGC\,6366 is 0.48~dex (see Table~\ref{tab:abund} for a full list of chemical abundances and the standard deviations of each chemical abundance). Except for two outliers (stars 05 and 11) the individual [O/Fe] deviates less than 0.05~dex from the mean value. The oxygen lines measured were the forbidden [OI] 6300.31 and 6363.78~\AA. It was not possible to measure the 6363-\AA\ feature in star 11, but its O abundance result seems to not be affected by this (see discussion about the Na-O anticorrelation in Section~\ref{subsec:zodd}).

\begin{figure}
	\includegraphics[width=\columnwidth]{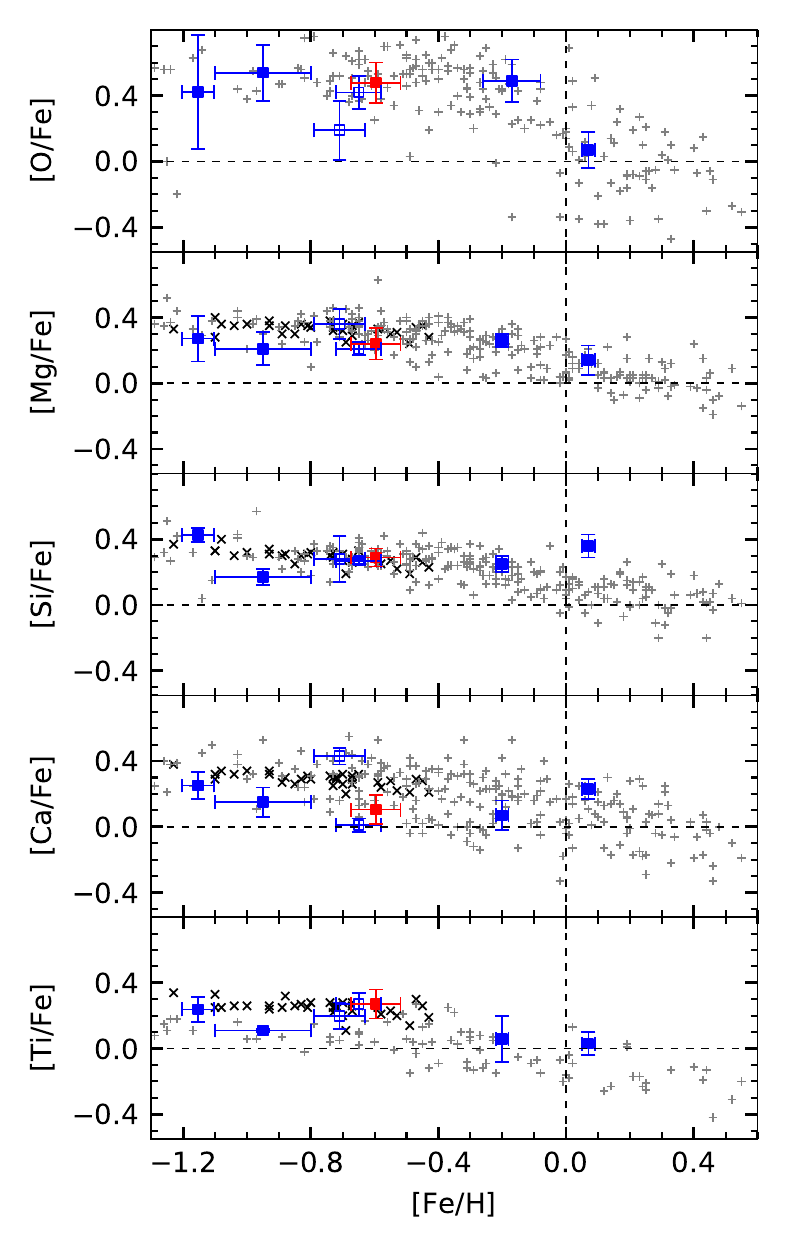}
    \caption{Comparsion of NGC\,6366 $\alpha$-elements abundance ratios with those of inner halo GCs, bulge field and GCs stars.
    \emph{Red square}: this work.
    \emph{Blue filled squares}: bulge GCs according to~\citet{Bica16}. NGC\,6266~\citep{Yong14}, NGC\,6522~\citep{Barbuy14}, NGC\,6553~\citep{AlvesBrito06} and NGC\,6528~\citep{Carretta01}.
    \emph{Blue open squares}: metal-rich GCs with R$_{\mathrm{GC}} >$ 4.5 kpc (also according to~\citeauthor{Bica16}), M\,71~\citep{Ramirez02} and 47\,Tuc~\citep{AlvesBrito05}.
    \emph{Grey crosses}: bulge field stars~\citep{AlvesBrito10,Johnson14}.
    \emph{Black 'x'}: 'high-$\alpha$' halo dwarfs from~\citet{Nissen10}.
    The error bars represent the star-to-star spread (1-$\sigma$) of each element in a given GC. The dotted lines are the solar values. Both lists of GCs in this and other captions are ordered by [Fe/H].}
    \label{fig:light}
\end{figure}

For [Mg/Fe] we found a mean value of 0.24~dex, $\sim$\,0.15~dex more metal-poor than the bulge field plateau estimated by~\citet{McWilliam16}. Stars 03 and 07 lack the blue part of their spectra, thus only the 6318.71-\AA\ line was measured for all targets, while the 5711-\AA\ feature was measured for the other six stars, without star-to-star variation beyond mean uncertainties. The spread found in our sample overlaps those of both bulge and metal-rich halo field stars selected for comparison, as shown in Fig.~\ref{fig:light}.

With a longer line list for Si in comparison with Mg, we measured $\langle$[Si/Fe]$\rangle$ = 0.29 for our sample. The Si abundances found in this work agree with both bulge and metal-rich halo field abundances, being on the high-$\alpha$ plateau.

The mean [Ca/Fe] is 0.10~dex, with six stars having a [Ca/Fe] ratio compatible with the solar value, within the associated uncertainties. The star-to-star variations of the abundances are smaller than the uncertainties. The mean [Ca/Fe] is 0.20~dex below that found in~\citet{Johnson16} for NGC\,6366 which is comparable to the mean uncertainty for Ca in this work, 0.19~dex. Inspection of the [Ca/Fe]--[Fe/H] plot in Fig.~\ref{fig:light} shows that NGC\,6366 is deficient in Ca when compared to the halo and that it overlaps the bulge field abundances, but one should also note that the bulge field scatter of Ca in the plot is very large, $\sim$ 0.40~dex.

\ion{Ti}{I} and \ion{Ti}{II} lines were measured, with $\langle$[Ti I/Fe]$\rangle$ = 0.27 and $\langle$[Ti II/Fe]$\rangle$ = 0.34, differing by less than 0.10~dex. The resulting mean from both species is [Ti/Fe] = 0.31. None of the Ti ions has $\rho >$~1, and the bigger $\rho$ value for \ion{Ti}{II} arises from its smaller mean uncertainty. When comparing with~\citet{AlvesBrito10} bulge field giants, Ti abundances in NGC\,6366 are on the higher limit of the spread for its [Fe/H] range, very close to the values calculated for 47\,Tuc by~\citet{AlvesBrito05}. The Ti abundances from this work agree well with those from metal-rich halo field stars obtained by~\citet{Nissen10}, as seen in Fig.~\ref{fig:light}.

Our results for the $\alpha$ elements match those found by~\citeauthor{Johnson16} for NGC\,6366, with the abundance of Ca being a possible exception. The abundances match both bulge and halo field abundances selected for comparison, with, again, the exception of Ca, which deviates from the halo pattern. We found similar abundances to those of~\citet{Dias16} for the stars in common with their work. For star 11 the [Fe/H] and [Mg/Fe] values differ by $\sim$ 0.05~dex, and for star 18 they differ by $\sim$ 0.10~dex.

The $\alpha$-enhancements are consistent with those of other GCs, and NGC\,6366 lies on the high-$\alpha$ plateau, with [$\alpha$/Fe] = 0.28. Despite being metal-rich, the GC position in the [$\alpha$/Fe]--[Fe/H] space is indicative of a relatively fast star formation rate instead of any Type Ia SNe enrichment of the previous generation of stellar population~\citep[as discussed, e.g. in][]{McWilliam97}.

\subsection{Iron peak elements}

Compared with other atomic species, the Iron peak elements are less explored in the context of chemical evolution of GCs. These elements are produced in massive stars, in hydrostatic and explosive conditions, but also in Type Ia SNe. In Type II SNe, V, Co and Ni are generated from explosive nuclear burning in complete Si-burning regions, while Cr and Mn are mainly synthesised in less deep incomplete Si-burning regions~\citep{Nomoto13}. For Mn,~\citet{Cescutti08} suggest the use of metallicity dependent yields for both types of SNe to reproduce observations in different populations. However,~\citet{Kobayashi09} argue that the Mn trend is caused by delayed enrichment of Type Ia SNe. The contribution of Type Ia SNe enhances Mn in high metallicity regimes, and this element can be useful to track the different astrophysical sites of nucleosynthesis~\citep{Kobayashi06}.

As for the $\alpha$ elements, none of the Fe-peak elements analysed in this work spreads beyond mean uncertainties, i.e., their $\rho$ is $\leq$ 1. The mean [V/Fe] is 0.13~dex. \ion{V}{I} is known for its extreme sensitivity to temperature variations in cold giants~\citep[as discussed in, e.g.,][]{Gratton04}. This affected the estimation of the uncertainties ($\langle \sigma_{*} \rangle$ = 0.28, see Table~\ref{tab:sensitiv} for T$_{\mathrm{eff}}$ sensitivities), but it did not translate into any spread. V shows some enhancement for most of the stars in the Baade's Window sample from~\citet{McWilliam94}, but in most cases the abundances from~\citeauthor{McWilliam94} and from this work overlap if we consider the uncertainties.

\begin{figure}
	\includegraphics[width=\columnwidth]{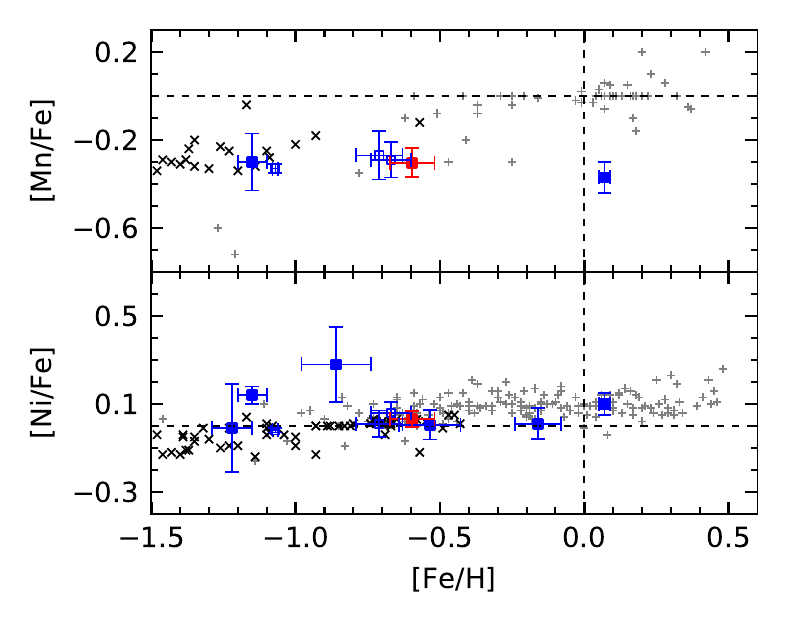}
    \caption{[Mn,Ni/Fe] vs. [Fe/H].
    The symbols are as in Fig.~\ref{fig:light}.
    \emph{bulge GCs}: for Mn and Ni, NGC\,6266~\citep{Yong14} and NGC\,6528~\citep{Carretta01}. For Ni, NGC\,6723~\citep{Gratton15}, Mercer 5~\citep{Penaloza15}, NGC\,6342~\citep{Johnson16} and NGC\,6553~\citep{Cohen99}.
    \emph{metal-rich GCs with R$_{\mathrm{GC}} >$ 4.5 kpc}: NGC\,6362~\citep{Massari17}, M\,71~\citep{Ramirez02} and 47\,Tuc~\citep{Carretta04}.
    \emph{Field}: bulge from~\citet{Johnson14} for Ni, and from~\citet{Barbuy13} for Mn. Inner halo field (with probability $>$0.9) from~\citet{Ishigaki13} for Mn and Ni and 'high-$\alpha$' halo dwarfs from~\citet[][Ni only]{Nissen10}.}
    \label{fig:fe_peak}
\end{figure}

For the odd-Z element Mn we chose to synthesise the \ion{Mn}{I} triplet at 6000~\AA. The mean [Mn/Fe] is $-$0.30~dex, similar to the mean [Mn/Fe] = $-$0.37 of a sample of 19 GCs analysed by~\citet{Sobeck06}. Inspecting Fig.~\ref{fig:fe_peak}, one can notice that NGC\,6366 falls within the abundance spread of bulge field stars, but GCs, in general, follow a constant trend, independent from field stars in higher metallicities. Other bulge GCs shown in the plot also have similar deficient [Mn/Fe], as do the most metal-rich ([Fe/H] $>$ $-$1.5) halo stars. NGC\,6366 behaves as a typical GC for Mn.

The [Cr/Fe] of NGC\,6366 is slightly below solar, with mean abundance of $-$0.03~dex, with $\sigma_{\mathrm{cl}}$ = 0.04 (N = 6 stars). For [Co/Fe], the mean is 0.08~dex, slightly above solar. As for Cr, Co abundances match those of the field~\citep[see e.g. Fig. 11 of][for comparison for both Cr and Co with the bulge field and~\citet{Nissen10} for Cr abundances in the halo field]{Johnson14}.

With several ($\sim$ 30) lines measured, our Ni measurements are solar. [Ni/Fe] has an average of 0.03~dex and the lowest star-to-star spread ($\rho$ = 0.16) of all the elements measured. In~\citet{Johnson16} the [Ni/Fe] of NGC\,6366 is 0.10~dex and the difference between both studies is essentially due to the solar abundances adopted. In the metallicity range shown in Fig.~\ref{fig:fe_peak}, the [Ni/Fe] values are between $-$0.20 and 0.20~dex for field stars, and most GCs shown fall in this interval -- with NGC\,6366 being very similar to GCs within its [Fe/H] region.

\subsection{Light odd-Z elements}
\label{subsec:zodd}

As far as the internal GC chemistry is concerned, the light odd-Z elements Na and Al are produced from proton-captures in the NeNa and MgAl cycles, respectively. The NeNa cycle occurs in the same site, hence the same temperature, where N is produced from O via the CNO cycle, while the temperature needed for the MgAl process is higher than the temperature required for Na production at the progenitor~\citep{Denisenkov90,Langer93,Salaris02}. As discussed in Section~\ref{sec:intro}, there is no consensus about the nature of the polluters which might generate the Na-O and Mg-Al anticorrelations found in many GCs.

For [Na/Fe] we found an average of 0.27~dex, with the largest spread for our sample ($\sigma_{\mathrm{cl}}$ = 0.15), comparable to the mean uncertainties (ratio $\rho$ = 1). This element has usually a large spread in GCs, and our results are similar to those of~\citet{Johnson16} for this GC. For Al, we have found an average abundance of [Al/Fe] = 0.36, an enhancement in agreement with the $\alpha$-like behaviour discussed by~\citet{McWilliam16}. In contrast with the possible [Na/Fe] variation, [Al/Fe] has $\rho$ = 0.38, thus no star-to-star variation was detected for this element.

\begin{figure}
	\includegraphics[width=\columnwidth]{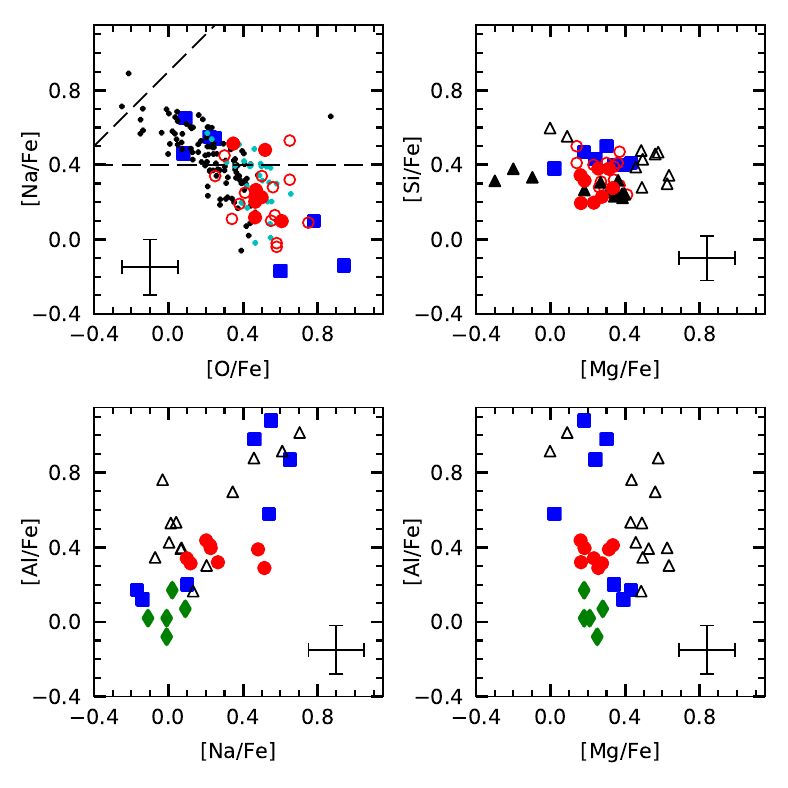}
    \caption{[X/Fe] abundance ratios for light elements.
    \emph{Red filled circles}: this work.
    \emph{Red open circles}: NGC\,6366 from~\citet{Johnson16}.
    \emph{Blue squares}: NGC\,6266~\citep{Yong14}.
    \emph{Green diamonds}: NGC\,6558~\citep{Barbuy07}.
    \emph{Black dots}: 47\,Tuc~\citep{AlvesBrito05,Carretta09b}.
    \emph{Cyan dots}: M\,71~\citep{Carretta09b}
    \emph{Black filled triangles}: NGC\,2808 and
    \emph{Black open triangles}: NGC\,7078~\citep{Carretta09c}
    The error bars are typical for this work.
    The dashed lines are the borders between P, I and E components (see text for discussion).}
    \label{fig:4plot}
\end{figure}

Fig.~\ref{fig:4plot} shows correlation/anticorrelation pairs of some of the light elements. The Na-O anticorrelation seems to be present in NGC\,6366, and its extension is small. In this context,~\citet{Carretta06} suggested the InterQuartile Range (IQR) as a quantitative measurement for the Na-O anticorrelation, because of its low sensitivity to outliers.~\citet{Carretta10} describes, among other global parameters relations, an anticorrelated relation between the IQR of [O/Na] and the absolute magnitudes of the GCs, a proxy measurement for their masses. According to~\citet[][2010 edition]{Harris96}, NGC\,6366 is one of the faintest GCs, with M$_{V}$ = $-$5.74. Two other GCs present in the Na vs. O plot, NGC\,6266 and 47\,Tuc, are among the most luminous Galactic GCs, and they have more extended anticorrelations. The IQR of NGC\,6366 in [O/Na] is 0.13~dex with our eight-star sample, while, e.g.,~\citet{Yong14} calculated 1.11 for NGC\,6266. Using data from the other high-resolution study of NGC\,6366~\citep{Johnson16}, we calculated an IQR of 0.28~dex for [O/Na] for a sample of 13 stars. M\,71 (NGC\,6838), another faint GC (M$_{V}$ = $-$5.61), which is also shown in Fig.~\ref{fig:4plot}, has [O/Na] IQR of 0.26.~\citeauthor{Carretta10} argue from evidence given in~\citet{Elsner08} that M\,71 lost a higher-than-average fraction of its original mass, and estimate a 'true' M$_V$ of $-$7.2 mag. Since both NGC\,6366 and M\,71 share similar values for mass, IQR of Na-O anticorrelation, metallicity and evidence of strong mass loss in the past (as mentioned for NGC\,6366 in Section~\ref{sec:intro}), we can take M\,71 as a comparative GC.

In our sample, only two stars -- 08 and 11 -- are above the border of the P ('primordial') and I ('intermediary') components, as defined in~\citet{Carretta09b}. Of those, only star 11 has subsolar [O/Na] = $-$0.16~dex, but still far from the threshold that characterizes the E ('extreme') population, [O/Na] $<$ $-$0.90. The border between P and I components, which varies with metallicity~\citep[e.g.][]{Carretta10}, was assumed at [Na/Fe] = 0.4 in NGC\,6366, the same value given for 47\,Tuc, since both GCs have similar [Fe/H]. Membership of the E component can be ruled out for all the stars by visual inspection of Fig.~\ref{fig:4plot}. The aforementioned stars 08 and 11 are inside the I component, but their (1-$\sigma$) error bars cross the P/I border. If we adopt a criterion that the measurements need to be located at more than 1-$\sigma$ from the border between components for a \emph{bona fide} characterization, star 18 also does not have a defined component and star 06 barely passes the requirement for the primordial population.

If we ignore the error bars, the fractions of P and I components found in this work are 0.75 and 0.25, respectively. However, no relations between GC parameters outlined in~\citet[][Figs. 11 to 16]{Carretta10} involving the P/I/E components can reproduce our results. Our values of [(Mg+Al+Si)/Fe] = 0.30 and [Ca/Fe] = 0.10 would give a P/I ratio $\geq$ 1, but also a nonzero fraction of E stars. The (low) Na-O IQR found in this work matches the absence of the extreme component, but fails to match with the (high) P and (low) I fractions, since, according to~\citeauthor{Carretta10}, the Na-O IQR is correlated with the former and anticorrelated with the later. If we use M\,71 as a proxy, the P/I/E fractions should be $\sim$ 0.28/0.72/0.00~\citep{Carretta09b}, i.e., the inverse found in our sample. However, for these P/I/E proportions, [Ca/Fe] should be at least 0.25~dex ($\sim$ 1$\sigma$ higher, but similar to the value found in~\citet{Johnson16}), and [(Mg+Al+Si)/Fe] should be at least 0.15~dex above the value derived in this work. Adopting the 1-$\sigma$ criterion, half of our sample is primordial, still a high fraction \emph{if the sample is representative of the entire GC}, which might not be the case because it has only eight stars. The Na-O anticorrelation of NGC\,6366 should be more extended if the GC has the estimated P/I ratio, and an extreme population should be present (see Fig. 11 of~\citet{Carretta10}), contradicting what is shown in Fig.~\ref{fig:4plot}.

Models of GC formation predict a homogeneous radial distribution after relaxation~\citep{Dercole08,Decressin10}, with the SG being more concentrated in the centre of the GC in the earlier stages of dynamical evolution~\citep{Vesperini13}. Indeed, some GCs show a spatial correlation between the predominance of each stellar population and the distance to the centre of the GC~\citep[e.g.][]{Bellini09,Zoccali09}, but~\citet{Larsen15} have found an inverse trend in M\,15, where giants of the first generation (FG) are more concentrated in the centre. Using the ratio between age and current half-mass relaxation time $t/t_{rh}$ to estimate the dynamical evolution of a GC, as suggested by~\citeauthor{Vesperini13}, we find $t/t_{rh} \approx$ 20 using the age from~\citet{Campos13} and $t_{rh}$ from~\citet[][2010 edition]{Harris96}. This value represents the stage when FG and SG should become well mixed, according to the Fig. 6 of~\citeauthor{Vesperini13}. Therefore, no bias in the P/I/E (FG/SG) fractions due to a narrow radial distribution of the sample should be expected. Also, neither our sample nor that of~\citet{Johnson16}, which has a broader radial distribution, present any correlation between star positions and the [O/Na] ratio.

As noted by~\citet{Charbonnel16}, the presence of multiple populations in star clusters is correlated with the high compactness index C$_{5} \equiv$ (M$_{cl}$/10$^{5}$ M$_{\sun}$)/(r$_{h}$/pc), where M$_{cl}$ is the cluster mass and r$_{h}$ is the half-mass radius. Gas expulsion tend to be more difficult in more compact star clusters due to their deeper potential well. NGC\,6366 is one of the least compact Galactic GCs, with C$_{5} =$ 0.058~\citep[][see also Fig. 2 of~\citeauthor{Charbonnel16}]{Krause16} for an estimated half-mass radius of 5.05 pc ($\approx$ 5~arcmin). As the small [O/Na] IQR of NGC\,6366 is well established from two different works, further studies of light elements with bigger samples and greater S/N ratio would be interesting to determine if NGC\,6366 really lacks a second generation (SG) of stars. This is apparently the case, as suggested by chemical analyses done to date (this work;~\citet{Johnson16}) and to the GC global parameters such as mass -- the least massive GC with clear FG/SG separation is NGC\,6362~\citep{Mucciarelli16}, which has M$_V$ = $-$6.95.

As previously mentioned, this work presents the first determination of Al abundances in NGC\,6366 using high-resolution spectra. While some GCs present a correlation between Na and Al~\citep[e.g.][]{Carretta09c}, inspection of Fig.~\ref{fig:4plot} reveals no such trend in NGC\,6366. The absence of correlation is related to the lack of variation of Al, a common feature in GCs more metal-rich than [Fe/H] $\geq$ $-$1.25 and also in the least massive ones (see, e.g,~\citet{Carretta09c} and~\citet{Cordero15}). Comparing NGC\,6366 with other GCs of similar metallicity, no Na-Al correlation is apparent in, e.g., M\,71~\citep{Cordero15} and the situation for 47\,Tuc is unclear~\citep[see, e.g.,~\citet{Carretta13} and][for different results]{Cordero14}, but we should point out that 47\,Tuc is a massive GC instead. As discussed in~\citet{Cordero15}, such dependence in the star-to-star variation of Al with metallicity can be reproduced by the AGB pollution scenario, as shown in~\citet[][Fig. 2]{Ventura14}, which reproduces to a rough approximation the [Al/Fe] spread in NGC\,6366 and its median value and also the greater scatter in the more metal-poor regime. However, we should note that the derived abundances in their models are for SG/Na-enhanced stars, whose existence in NGC\,6366 is not clear. In addition, leakage of the MgAl process in Si, which is found in some GCs~\citep[see discussion in][and references therein]{Gratton12}, is not found for NGC\,6366, in contrast with GCs like NGC\,2808 and NGC\,7078, as seen in the top right plot of Fig.~\ref{fig:4plot}.

\subsection{Neutron-capture elements}

Elements with atomic mass A $>$ 56 are synthesised by capture of neutrons by Fe-peak nuclei in a diversity of astrophysical sites~\citep{Karakas14}. The main channels of neutron capture are the s-process and the r-process. The s-process ('slow') is characterized by neutron-capture occurring on a time-scale longer than that of $\beta$-decay, i.e., a $\beta$-emission will occur before the capture of another neutron by the nucleus. It is done in low neutron density environments -- N$_{n} \leqslant$ 10$^{8}$~neutrons~cm$^{-1}$~\citep{Karakas14}. The He-burning core and convective C-burning shell (atomic mass up to A $\sim$ 90) of massive stars~\citep[e.g.][]{Pignatari10,Frischknecht12,Frischknecht16} and interiors of AGB stars (atomic mass above A $\sim$ 90)~\citep[e.g.][]{Bisterzo11,Bisterzo14,Bisterzo17} are the main sites. In the solar neighbourhood, elements such as Y and Ba can be used as tracers of the s-process~\citep{Simmerer04}. The r-process ('rapid') differs from the s-process by having a rate of neutron-capture faster than the time-scale of $\beta$-decay. Despite Type II SNe being traditionally suggested as the main site of this process, its place of origin is not well understood~\citep{Thielemann11}. Other proposed sites are collisions between neutron stars and black holes~\citep{Lattimer74} and neutron star mergers~\citep[e.g.][]{Eichler89,Freiburghaus99,Rosswog14}. Recent observations give evidence to the collision/merger scenarios~\citep{Tanvir13,Drout17}. Observationally, it is well known that Eu is a good tracer of the r-process in the solar neighbourhood~\citep{Simmerer04}.

\begin{figure}
	\includegraphics[width=\columnwidth]{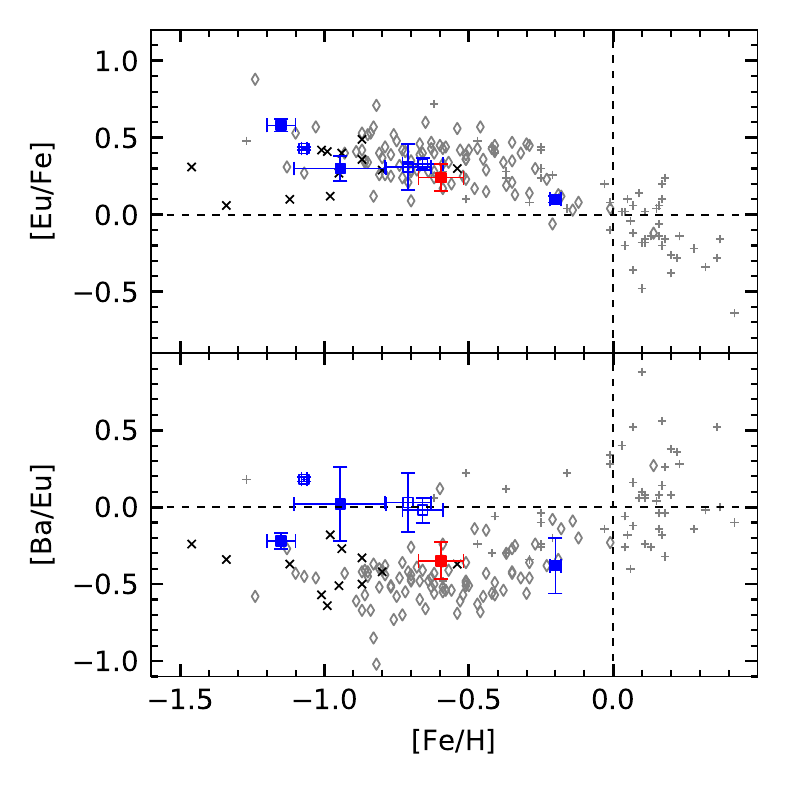}
    \caption{\emph{Top}: [Eu/Fe] vs. [Fe/H].
    	\emph{Bottom}: [Ba/Eu] vs. [Fe/H].
        The symbols are as in Fig.~\ref{fig:light}, with grey open diamonds representing thick disc field stars.
        \emph{Bulge GCs}: NGC\,6266~\citep{Yong14}, NGC\,6522~\citep{Barbuy14}, NGC\,6553~\citep{AlvesBrito06}.
        \emph{Metal-rich GCs with R$_{GC} >$ 4.5 kpc}: NGC\,6362~\citep{Massari17}, M\,71~\citep{Ramirez02}, 47\,Tuc~\citep{AlvesBrito05}.
        \emph{Field}: bulge stars from~\citet{Swaelmen16}, thick disc and halo stars from~\citet{Reddy06}.
        The error bars represent the standard deviations of the GCs samples. The dotted lines are the solar values.}
    \label{fig:ncap}
\end{figure}

The lightest neutron-capture element measured in this work, Y, has a mean abundance of [Y/Fe] = $-$0.18. It was not possible to infer Y abundances for stars 03 and 07 because there were no lines available in their spectra. The mean Y abundance is similar to those of NGC\,6388~\citep{Carretta07}, NGC\,6441~\citep{Gratton06} and M\,71~\citep{Ramirez02}, all of them in the range $-$0.80 $<$ [Fe/H] $<$ $-$0.40. The more metal-poor GCs NGC\,6266 and NGC\,6522 present enhanced [Y/Fe] values~\citep[][and~\citet{Barbuy14}, respectively]{Yong14}. In the (very) metal-poor GC NGC\,6397,~\citet{Lind11} found [Y/Fe] = $-$0.26 using the same \ion{Y}{II} lines (5087 and 5200~\AA) from this work. NGC\,6397 was also studied by~\citet{James04}, with Y abundances similar to those of~\citeauthor{Lind11}, and their mean [Y/Fe] for 47\,Tuc subgiants is $-$0.11~dex. However, the results of this element for 47\,Tuc must be interpreted with caution, since its [Y/Fe] found in the literature range from $-$0.13 to 0.65~dex~\citep[e.g.,][]{Wylie06,McWilliam08}. The differences could be due to different methods and line lists employed. \citet{Thygesen14} attributes the possible overestimation of abundances in \ion{Y}{II} lines in some studies as due to their choice of not applying hyperfine structure treatment. In spite of that, the agreement between the works of~\citeauthor{James04} and ~\citeauthor{Lind11} using different line lists gives some confidence to our result, derived from the same lines used by~\citeauthor{Lind11}.

For Ba, a mean abundance of $-$0.11~dex is found, with star-to-star spread of $\sigma$ = 0.09. Most stars are less than 1-$\sigma$ from solar [Ba/Fe]. The line used to derive Ba abundances was the 5853-\AA\ feature, whose hyperfine and isotopic splitting is negligible~\citep[see, e.g.,][]{Mashonkina06}. The result deviates from the raising trend of [Ba/Fe] with [Fe/H] in Galactic GCs identified by~\citet{Dorazi10}, which analysed a homogeneous sample of more than 1200 GC stars. In their analysis, [Ba/Fe] reaches a peak of $\sim$ 0.60 at [Fe/H] $\sim$ $-$1.10, while the most metal-rich GCs of their sample, 47\,Tuc and M\,71, seem to deviate, being less Ba-abundant than expected by a linear fit (see their Fig. 1). The [Ba/Fe] derived for NGC\,6366 deviates even more from the raising trend, and falls in the [Ba/Fe] range of the field stars with the same metallicity~\citep[see also Fig. 6 of][]{Swaelmen16}. It should be noted that, except for the aforementioned 47\,Tuc and M\,71, all GCs from~\citeauthor{Dorazi10} have [Fe/H] $<$ $-$1. \citet{Barbuy14} and~\citet{Yong14} measured enhanced [Ba/Fe] $\sim$ 0.35 for the intermediate-metallicity GCs NGC\,6522 and NGC\,6266, respectively, with some agreement with the trend from~\citeauthor{Dorazi10}. In the metal-rich range,~\citet{AlvesBrito06} found that NGC\,6553 is depleted in this element, with mean [Ba/Fe] = $-$0.28, while~\citet{Carretta01} measured a mild enhancement of 0.14~dex for NGC\,6528. As for Y, NGC\,6366 falls in the group of GCs which are deficient in Ba.

The abundance of r-process elements in Galactic GCs is enhanced w.r.t. to the Sun, with a few exceptions~\citep[see, e.g.,][]{Gratton04}. For NGC\,6366 we calculated $\langle$[Eu/Fe]$\rangle$ = 0.24 ($\sigma$ = 0.09), assuming solar isotopic ratios of \ce{^{151}Eu} and \ce{^{153}Eu}~\citep[][Table 3]{Asplund09}. As seen in Fig.~\ref{fig:ncap}, the mean [Eu/Fe] derived in this work lies in the bulge/thick-disc field range. Also, as the mean values of [Eu/Fe] and [$\alpha$/Fe] are very similar in our sample, we suggest that, taking the uncertainties into account, both $\alpha$ and r-process elements in NGC\,6366 might have similar astrophysical origin.

Since neutron-capture elements are produced by both s- and r- processes, it can be useful to calculate the ratio [s/r] between the abundance of elements whose production is dominated, respectively, by the s-process and the r-process. A higher (positive) [s/r] indicates dominance of nucleosynthesis of heavy elements in sites associated with the s-process, e.g, AGB stars; while a lower (negative) [s/r] suggests predominance of sites associated with the r-process, e.g., Type II SNe and/or neutron star mergers. We chose to use the Ba results as a proxy to the s-process history due to its higher s-process proportion in the solar system than Y, while Eu is the r-process representative. For NGC\,6366 we found $\langle$[Ba/Eu]$\rangle$ = $-$0.35. This ratio suggests some dominance of r-process in the formation of heavy elements in this GC, since it is closer to the pure r-process solar ratio [Ba/Eu] = $-$0.82 than to the pure s-process solar ratio [Ba/Eu] = 1.60~\citep[both values derived from Table 1 of][]{Sneden08}. The [Ba/Eu] ratio also lies in the (mostly thick disc) field range, and is lower than those of the GCs of similar metallicities evaluated in Fig.~\ref{fig:ncap}. Summing up the [Ba/Eu] value and the similarities between the [Eu/Fe] and [$\alpha$/Fe] ratios, the primordial composition of NGC\,6366 may have been previously enriched by objects with short time-scale, likely Type II SNe, with mild contribution of other sites like AGBs.

\subsection{Global abundance pattern}

\begin{figure}
	\includegraphics[width=\columnwidth]{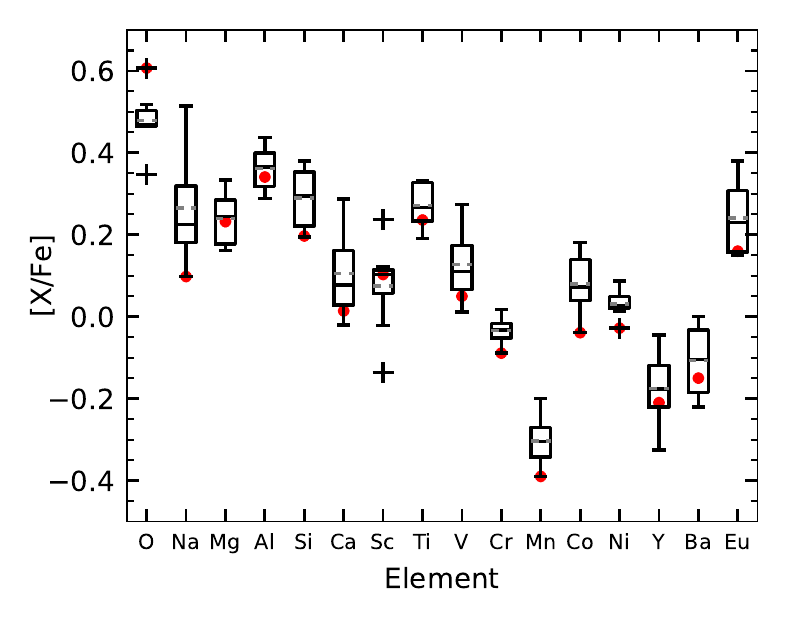}
    \caption{Star-to-star abundance spread for all elements measured in this work, except Fe. \emph{Boxes}: the interquartile ranges (IQR). \emph{Whiskers}: data inside 1.5*IQR limits. \emph{Black horizontal lines}: medians. \emph{Grey dashed lines}: means. \emph{Crosses}: outliers. \emph{Red circles}: abundances of star 05.}
    \label{fig:boxplot}
\end{figure}

We present in Fig.~\ref{fig:boxplot} the global abundance pattern for NGC\,6366, with star 05 highlighted. As previously discussed in Section~\ref{sec:kin},~\citet{Sariya15} gave zero membership probability to star 05 based on its proper motion, while our radial velocity measurements suggest a value compatible with NGC\,6366 kinematics. Chemically, this star follows the general abundance pattern of the GC. Despite being the less abundant star in some (but not all) Fe-peak elements, its deviation is negligible when compared to the uncertainties and does not contribute to any detectable star-to-star abundance spread. Also, star 05 presents a low-Na/high-O pattern that is typical of GCs stars.

Past studies noted similarities between NGC\,6366 and 47\,Tuc, such as their HB morphology, RR Lyrae population and metallicity~\citep[e.g.][and references therein]{Sarajedini07}. However, 47\,Tuc is one of the most massive Galactic GCs, as opposed to NGC\,6366, and it may have had a different dynamical history. Indeed, it has $t/t_{rh} \approx$ 3~\cite[][2010 edition,~\citet{Cordero15}]{Harris96}, much less than the estimation given for NGC\,6366 in Sect.~\ref{subsec:zodd}. Chemically, 47\,Tuc, unlike NGC\,6366, has a clear Na-O anticorrelation -- even with the anticorrelation extension in 47\,Tuc being smaller than what should be expected for its mass, the Na-O anticorrelation is present in this GC. Also, as discussed in Sect.~\ref{subsec:zodd}, it is not clear if its star-to-star variation (or lack of thereof) in Al is similar to the result found in this work for NGC\,6366. Of the s-process dominated elements evaluated here, 47\,Tuc presents enhancement, while we show in this work that NGC\,6366 is deficient -- despite the Y measurements variation in 47\,Tuc, as discussed in the previous section, its reported [Ba/Fe] values are more consistent~\citep[e.g.][]{AlvesBrito05,Dorazi10,Thygesen14}.

On the other hand, NGC\,6366 and M\,71 share similar global parameters, as discussed in Sect.~\ref{subsec:zodd}. Chemically, both GCs have similar Na-O anticorrelations (see Fig.~\ref{fig:4plot}) and both lack star-to-star variation in Al. NGC\,6366 and M\,71 abundances only differ more than $\sim$ 0.15~dex in Ba, and possibly Ca -- however, the [Ca/Fe] from~\citet{Ramirez02} shown in Fig.~\ref{fig:light} may be overestimated, see e.g. the results of~\citet{Melendez09} and~\citet{Meszaros15} for M\,71. Despite the difference in Ba, the [Y/Fe] values of both clusters match, but a detailed comparison of both GCs in the s-process family would clarify their level of similarity. They possibly share dynamical histories alike, due to their masses and past mass loss. In this sense, a detailed analysis of NGC\,6366 orbit, like that done to M\,71 and other GCs in~\citet{Dinescu99} would be useful. Judging by their chemical resemblance, NGC\,6366 must have a low-eccentricity, low-inclination orbit, occupying the outer bulge/inner halo of the Milky Way. On the other hand, an orbit with higher eccentricity and low perigalactic distance could be a hint to a bulge origin.

In summary, NGC\,6366 has a bulge-like metallicity, while both its kinematics and chemical similarity with M\,71 and, possibly, 47\,Tuc indicate otherwise. Also, a further homogeneous chemical analysis including NGC\,6366 and these two GCs could help to improve our understanding of their internal chemistry and, maybe, clarify the situation of NGC\,6366 in the Galactic context.

\section{Conclusions}
\label{sec:conclusion}

Our analysis confirms NGC\,6366 as a metal-rich GC, with no star-to-star [Fe/H] spread, based on eight giant stars. None of the elements measured present star-to-star variation greater than the uncertainties, i.e., the metallic content of NGC\,6366 is very homogeneous.

Being considered a GC with halo kinematics, NGC\,6366 has a chemical abundance pattern which can be consistent with both the bulge and the $\alpha$-rich, metal-rich halo. The [Fe/H] abundance of $-$0.60~dex puts this GC in the metal-rich group of GCs, dominated by GCs associated with the bulge. However, it also presents similarities with the non-bulge GC M\,71, which orbits the Galaxy in the inner thick disc according to~\citet{Dinescu99}. Given these similarities, NGC\,6366 possibly has an orbit with low eccentricity close to the Galactic plane, but a detailed study of its orbital dynamics, preferentially with Gaia proper motion data, should be made before drawing conclusions. A resemblance with 47\,Tuc is not evident chemically, but additional data from Al and the s-process elements in both GCs may be useful to clarify this point.

On the question of multiple populations, we have not found any unambiguous SG star. Based on the results of light odd-Z abundances, it can be assumed that our entire sample, or at least most of it, is from the FG. When comparing this work with the analysis of~\citet{Johnson16}, both samples overlap in Na-O diagram, showing a barely detectable, if any, Na-O anticorrelation. Our study found no Mg-Al anticorrelation and no star-to-star variation in Al.

If NGC\,6366 contains a simple stellar population, which is possibly the case, it enters the list of Galactic GCs without detection of multiple populations, adding evidence that, at least in the Galaxy, such phenomenon occurs only in GCs brighter than M$_{V} \approx$ $-$7. Additional high-precision photometric data as well as studies focused upon high-resolution spectroscopy at optical and infrared wavelengths targeting larger samples and fainter magnitudes would help to improve our understanding of this very reddened GC.

\section*{Acknowledgements}

We thank the anonymous referee for important comments and suggestions. Partial financial support for this research comes from BIC-UFRGS institutional program and CNPq (Brazil). This publication makes use of data products from the Two Micron All Sky Survey, which is a joint project of the University of Massachusetts and the Infrared Processing and Analysis Center/California Institute of Technology, funded by the National Aeronautics and Space Administration and the National Science Foundation. The Digitized Sky Surveys were produced at the Space Telescope Science Institute under U.S. Government grant NAG W-2166. The images of these surveys are based on photographic data obtained using the Oschin Schmidt Telescope on Palomar Mountain and the UK Schmidt Telescope. {\sc iraf} is distributed by the National Optical Astronomy Observatories, which are operated by the Association of Universities for Research in Astronomy, Inc., under cooperative agreement with the National Science Foundation. Based on observations collected at the European Organisation for Astronomical Research in the Southern Hemisphere under ESO programme 69.B-0467. Based on observations obtained at the Canada-France-Hawaii Telescope (CFHT) which is operated by the National Research Council of Canada, the Institut National des Sciences de l´Univers of the Centre National de la Recherche Scientique of France, and the University of Hawaii.

\bibliographystyle{mnras}
\bibliography{refer}

\appendix

\section{Tables and figures}

\begin{table*}
\caption{Uncertainties, in~dex, for the elements measured by the equivalent widths method.}
\label{tab:unc}
\begin{tabular}{lrrrrrrrr}
\hline
Element     &   01 &   03 &   05 &   06 &   07 &   08 &   11 &   18 \\
\hline
\ion{Fe}{I} & 0.19 & 0.19 & 0.21 & 0.19 & 0.20 & 0.19 & 0.19 & 0.20 \\
\ion{Fe}{II}& 0.19 & 0.28 & 0.22 & 0.21 & 0.19 & 0.20 & 0.28 & 0.29 \\
\ion{O}{I}  & 0.14 & 0.14 & 0.18 & 0.16 & 0.16 & 0.14 & 0.14 & 0.16 \\
\ion{Na}{I} & 0.15 & 0.13 & 0.13 & 0.13 & 0.13 & 0.15 & 0.18 & 0.19 \\
\ion{Mg}{I} & 0.15 & 0.07 & 0.17 & 0.13 & 0.09 & 0.11 & 0.24 & 0.24 \\
\ion{Al}{I} & 0.15 & 0.11 & 0.12 & 0.11 & 0.13 & 0.09 & 0.22 & 0.12 \\
\ion{Si}{I} & 0.14 & 0.15 & 0.10 & 0.07 & 0.11 & 0.08 & 0.13 & 0.13 \\
\ion{Ca}{I} & 0.18 & 0.20 & 0.18 & 0.19 & 0.19 & 0.16 & 0.21 & 0.20 \\
\ion{Sc}{II}& 0.16 & 0.15 & 0.18 & 0.15 & 0.15 & 0.19 & 0.18 & 0.19 \\
\ion{Ti}{I} & 0.24 & 0.24 & 0.24 & 0.22 & 0.22 & 0.22 & 0.28 & 0.26 \\
\ion{Ti}{II}& 0.16 & 0.15 & 0.19 & 0.19 & 0.14 & 0.19 & 0.19 & 0.21 \\
\ion{V}{I}  & 0.27 & 0.28 & 0.25 & 0.23 & 0.29 & 0.27 & 0.33 & 0.35 \\
\ion{Cr}{I} & 0.19 &$\cdots$ & 0.19 & 0.26 &$\cdots$ & 0.18 & 0.28 & 0.19 \\
\ion{Co}{I} & 0.15 & 0.17 & 0.16 & 0.15 & 0.15 & 0.17 & 0.14 & 0.15 \\
\ion{Ni}{I} & 0.20 & 0.17 & 0.19 & 0.19 & 0.19 & 0.17 & 0.19 & 0.21 \\
\hline
\end{tabular}
\end{table*}

\begin{figure*}
	\includegraphics[width=\columnwidth]{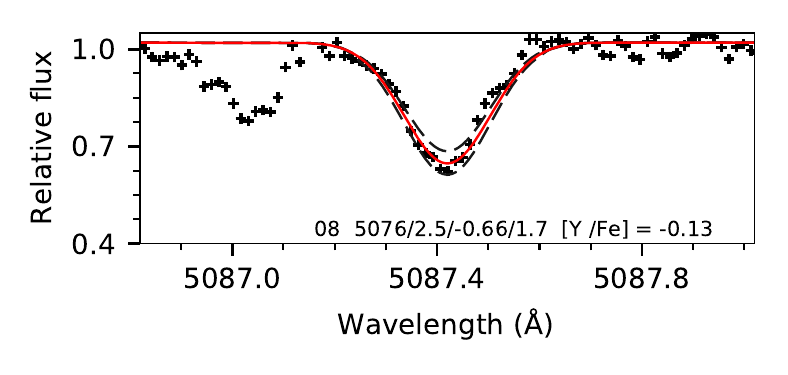}
    \caption{Example of spectral synthesis of the 5087~\AA\ line of Y. The symbols are as given in Fig.~\ref{fig:synth_Mn}, except that only 1-$\sigma$ deviations are shown.}
    \label{fig:synth_Y}
\end{figure*}

\begin{figure}
	\includegraphics[width=\columnwidth]{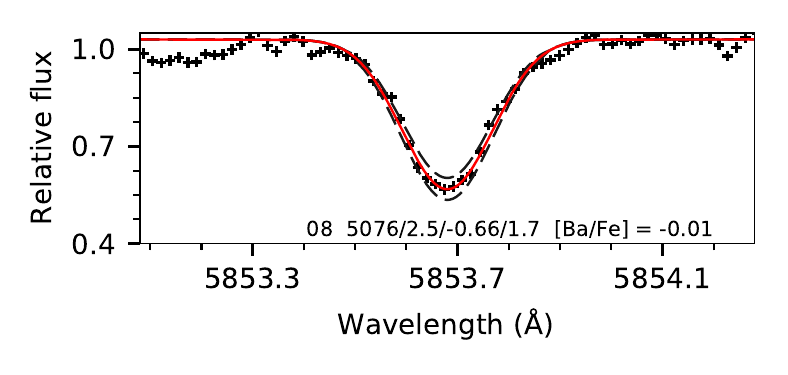}
    \caption{Example of spectral synthesis of the 5853~\AA\ line of Ba. The symbols are as given in from Fig.~\ref{fig:synth_Y}.}
    \label{fig:synth_Ba}
\end{figure}

\begin{figure}
	\includegraphics[width=\columnwidth]{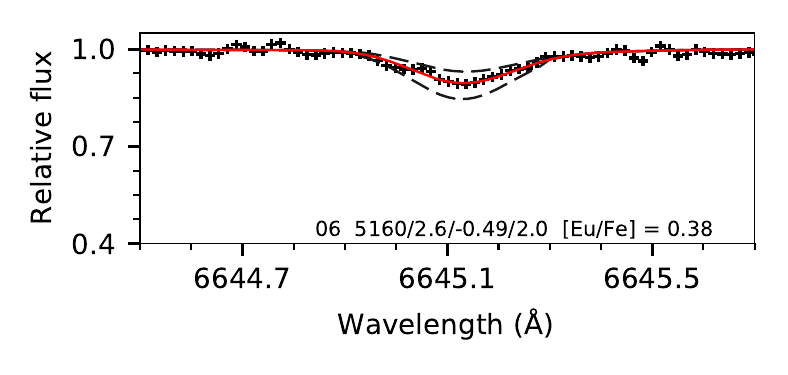}
    \caption{Example of spectral synthesis of the 6645~\AA\ line of Eu. The symbols are as given in Fig.~\ref{fig:synth_Y}.}
    \label{fig:synth_Eu}
\end{figure}

\bsp	
\label{lastpage}
\end{document}